\documentclass{aa}  

\usepackage{graphicx}
\usepackage{txfonts}
\usepackage{xcolor}
\usepackage{mathrsfs}
\usepackage{amssymb,amsmath}
\usepackage{newtxtext,newtxmath}
\usepackage{amsmath}    
\usepackage{amssymb}    
\DeclareMathAlphabet{\mathbi}{OT1}{ptm}{bx}{it}
\SetMathAlphabet\mathbi{bold}{OT1}{ptm}{bx}{it}
\newcommand{\bd}[1]{\mbox{\boldmath $#1$}}
\usepackage{hyperref}
\hypersetup{
    colorlinks=true,
    citecolor=blue,
    linkcolor=blue,
    filecolor=magenta,      
    urlcolor=cyan,
}
\begin{document}

   \title{Disentangling the optical AGN and host-galaxy luminosity with a probabilistic flux variation gradient\thanks{The PFVG code can be downloaded from: \url{https://github.com/HITS-AIN/ProbabilisticFluxVariationGradient.jl/}} \fnmsep \thanks{Instructions and specific examples used in this paper can be found in: \url{https://github.com/HITS-AIN/PFVG_AA2021}}}
   
   \titlerunning{Probabilistic flux variation gradient}


   \author{N. Gianniotis\inst{1}
          \and
          F. Pozo Nu\~nez \inst{1}
          \and
          K. L. Polsterer\inst{1}
          }

   \institute{Astroinformatics, Heidelberg Institute for Theoretical Studies, Schloss-Wolfsbrunnenweg 35, 69118 Heidelberg, Germany\\
              \email{nikos.gianniotis@h-its.org; francisco.pozonunez@h-its.org}}

   \date{Received ??, 2021; accepted ??, 2021}

 \abstract{We present a novel probabilistic flux variation gradient (PFVG) approach to separate the contributions of active galactic nuclei (AGN) and host galaxies in the context of photometric reverberation mapping (PRM) of AGN.} {We explored the ability of recovering the fractional contribution in a model-independent way using the entire set of light curves obtained through different filters and photometric apertures simultaneously.} {The method is based on the observed "\textup{}bluer when brighter" phenomenon that is attributed to the superimposition of a two-component structure; the red host galaxy, which is constant in time, and the varying blue AGN. We describe the PFVG mathematical formalism and demonstrate its performance using simulated light curves and available PRM observations.} {The new probabilistic approach is able to recover host-galaxy fluxes to within 1\% precision as long as the light curves do not show a significant contribution from time delays. This represents a significant improvement with respect to previous applications of the traditional FVG method to PRM data.} {The proposed PFVG provides an efficient and accurate way to separate the AGN and host-galaxy luminosities
 in PRM monitoring data. The method will be especially helpful in the case of large upcoming photometric survey telescopes such as the public optical/near-infrared Legacy Survey of Space and Time (LSST) at the Vera C. Rubin Observatory. Finally, we have made the algorithms freely available as part of our \textit{\textup{Julia}} PFVG package.}

   \keywords{galaxies: active --galaxies: quasars
          --galaxies: nuclei --galaxies: Seyfert
          --methods: statistical --methods: numerical
               }

   \maketitle
%

\section{Introduction}

Powered by supermassive black holes in their centers, active galactic nuclei (AGN) are considered to be the most energetic sources in the Universe. The involved physical regions are extremely compact and hard to resolve with existing instrumentation. As a result,  AGN often appear as a point-like sources compared with their host galaxy, which can have a great variety of morphologies (e.g., \citealt{2009ApJ...691..705G}; \citealt{2014MNRAS.439.3342V}), and whose degree of contribution to the total flux varies significantly (e.g., \citealt{2010MNRAS.405..718P}; \citealt{2012A&A...541A.118P}; \citealt{2014MNRAS.440..476F}; \citealt{2015MNRAS.454.4103B}). 

The case of local Seyfert galaxies is particularly challenging. In these galaxies, the starlight contribution can reach up to 50 to $\sim$80\% of the total optical (e.g., \citealt{2009ApJ...697..160B}; \citealt{2010ApJ...711..461S}; \citealt{2012A&A...545A..84P}; \textcolor{blue}{2013}) and infrared luminosities (e.g., \citealt{1998MNRAS.297...18G}; \citealt{2004MNRAS.350.1049G}; \citealt{2006ApJ...639...46S}; \citealt{2014A&A...561L...8P}; \textcolor{blue}{2015}; \citealt{2014ApJ...788..159K}). 
An overestimation of the AGN luminosity due to the contamination of its host galaxy has strong implications on the refinement of scaling relations for single-epoch black hole estimates, such as the relation of the  H$\beta$ broad-line region (BLR) size to the $5100$\,\AA\ monochromatic luminosity ($R_{\rm BLR}\propto L_{\rm{AD}}^{\alpha}$; \citealt{2013ApJ...767..149B}). The $R_{\rm BLR}\propto L_{\rm{AD}}^{\alpha}$ relation offers the possibility of using quasars as standard candles for cosmological distance studies (\citealt{1999OAP....12...99O}; \citealt{2011ApJ...740L..49W}). However, this can only be achieved if the scatter is reduced considerably. The scatter in this relation appears not only because of the noise in time-delay measurements, from which BLR sizes are inferred, but also from erroneous separation of AGN and host galaxy luminosities. This has been demonstrated by \cite{2013ApJ...767..149B}, who obtained an improved slope of $\alpha = 0.533_{-0.033}^{+0.035}$ for a sample of 41 reverberation-mapped Seyfert 1 galaxies after correcting for starlight contamination.

To date, there are two main methods for isolating  nuclear flux: fitting galaxy templates to the observed AGN spectrum (e.g., \citealt{2015A&A...575A..22M}), and modeling the host galaxy profile using high-resolution images (e.g., \citealt{2002ApJ...569..624P}; \citealt{2009ApJ...697..160B}, \textcolor{blue}{2013}, using Hubble Space Telescope, HST, imaging). The first method can be directly applied to spectroscopic reverberation mapping (SRM) campaigns, while the second is mostly restricted to local Seyfert galaxies in which the host galaxy is optically resolved, and it depends strongly on the adopted galaxy model. However, neither method is  efficient in the case of large monitoring programs, in particular, in the case of large-scale photometric reverberation-mapping (PRM) surveys. This raises a practical concern regarding the processing of the unprecedented amount of data that will be provided by large ground-based photometric surveys (LSST\footnote{Legacy Survey of Space and Time. \url{https://www.lsst.org/}}; \citealt{2009arXiv0912.0201L}, SDSS\footnote{Sloan Digital Sky Survey. \url{https://www.sdss.org/future/}}-V; \citealt{2019BAAS...51g.274K}), and satellite missions (TESS\footnote{Transiting Exoplanet Survey Satellite. \url{https://tess.mit.edu/}} ; \citealt{2018RNAAS...2...47J}, JWST\footnote{James Webb Space Telescope. \url{https://www.jwst.nasa.gov/}}; \citealt{2017ApJ...838..117N}, Euclid\footnote{European Space Agency mission. \url{https://sci.esa.int/web/euclid}}; \citealt{2018LRR....21....2A}) in the near future.

An alternative approach that does not require the use of high spatial resolution images is the flux variation gradient method (FVG, \citealt{1992MNRAS.257..659W}; \citealt{1997MNRAS.292L..50G}; \citealt{1998MNRAS.297...18G}). The FVG is based on the flux-flux diagrams by \cite{1981AcA....31..293C}, who attributed the observed "\textup{bluer when brighter}" phenomenon to the superimposition of a two-component structure: the contribution of a red host galaxy that is constant in time, including nonvarying emission lines, and the varying contribution of an AGN with constant blue color. 

Recently, given a well-defined range of host-galaxy slopes 
(e.g., \citealt{2010ApJ...711..461S}), the FVG has been successfully employed to separate  host and AGN contributions in narrow- and broad-band PRM campaigns (e.g., \citealt{2012A&A...545A..84P}; \citealt{2015A&A...581A..93R}; \citealt{2019MNRAS.490.3936P}; \citealt{2019NatAs...3..251C}), as well as in dust near-infrared RM studies (e.g., \citealt{2014A&A...561L...8P}; \textcolor{blue}{2015}; \citealt{2015ApJ...801..127V}; \citealt{2018A&A...620A.137R}). Moreover, the FVG method can be applied without a prior assumption about the host-galaxy slopes. In this case, the method requires simultaneous observations with at least one photometric band in which the host-galaxy contribution is negligible (e.g., $\sim2000$\,\AA), thus providing a lower limit for the host-galaxy flux (\citealt{2017ApJ...835...65S}; \citealt{2018MNRAS.480.2881M};  \citealt{2020ApJ...896....1C}; \citealt{2020MNRAS.498.5399H}). However, the actual FVG application has considerable difficulties and limitations that have not yet been fully addressed and quantified. For instance, it is unclear whether an FVG analysis can give a reliable estimate of the host-galaxy contribution based solely on traditional linear regression analysis. It might equally rather return biased results given the observational errors in the data, different amplitudes, and variability features in the light curves. It is also unclear to which degree FVGs are affected by light curves with time delays obtained from PRM observations. To answer these questions in this paper, we study the reliability of the FVG method under different observational conditions, and introduce a new probabilistic approach that can be efficiently applied to PRM monitoring data.

\section{Methods}\label{sec2}

In the following section, we briefly revisit the FVG method and introduce a new probabilistic reformulation. We note that a review of the FVG method and its application to PRM data can be found in \cite{2012A&A...545A..84P}. Here we list the formulae and principle for the sake of comprehensiveness.

\subsection{Flux variation gradient}
\label{subsec2}
The purpose of the FVG is to separate the AGN and host-galaxy contributions given the observed total flux and a ratio of colors characterizing the host galaxy in question.
The observed total fluxes $y_i(t)$ are the sum of a host-galaxy flux that is constant in time, $g_i(t)=g_i,$ and the variable AGN flux $v_i(t)$,
\begin{eqnarray}
y_{i}(t) = v_i(t) + g_i \ .
\label{eq:flux_equals_ang_plus_galaxy}
\end{eqnarray}
Two assumptions underpin the FVG method.
The first assumption is that the AGN fluxes $v_i(t)$ form a line that goes through the origin. The second assumption, supported by observations, is that total fluxes $y_{i}(t)$ obtained through different photometric bands $i\in\{1,\dots,F\}$, where $F$ is the total number of filters, follow a linear relation.
The AGN optical spectral shape does not change, that is, the ratio $\frac{v_i(t)}{v_j(t)} = \Gamma_{ij}$ is a constant, known as
the flux variation gradient, as described by \cite{1992MNRAS.257..659W}.
Hence, we can write the assumption of a linear relation as 
\begin{eqnarray}
y_{i}(t) = \Gamma_{ij} y_{j}(t) + g_{i} - \Gamma_{ij} g_{j} \ .
\label{eq:fall_on_a_line}
\end{eqnarray}

\begin{figure}
 \includegraphics[width=\columnwidth]{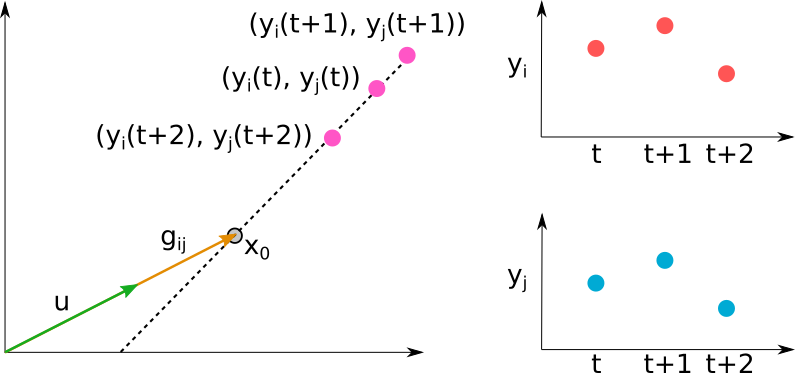}
 \caption{Sketch of FVG. Observations in filters $i$ and $j$ measured at three different time instances (right). By pairing flux values of co-occurring observations, we form points in the flux-flux plot that fall on a line (dashed). Vector $\bd{g}_{ij}$ (in brown) corresponds to the unobserved host-galaxy and defines a line $\bd{g}_{ij} \cdot x$ that intersects the dashed line at $\bd{x}_0$, the point of (unobserved) minimum AGN activity. The FVG method consists of finding the intersection of $\bd{u}\cdot x$ and the dashed line; $\bd{u}$ (in green) is a vector that has the same direction (i.e., same colors) as $\bd{g}_{ij}$ and therefore also intersects the dashed line at $\bd{x}_0$.}
 \label{fig:fvg_2D}
\end{figure}

Figure~\ref{fig:fvg_2D} sketches the FVG method.
The right side shows observations in two different filters. By pairing flux values of observations that occur at the same time, we form points $(y_i(t), y_j(t))$ in the flux-flux plot that fall on the dashed line, as dictated by Equation~\eqref{eq:fall_on_a_line}.
Each of the instantiated points $(y_i(t),y_j(t))$ in the flux-flux plot can be understood as a vector that results from adding the unobserved AGN vector $(v_i(t), v_j(t))$ (not plotted in Fig. \ref{fig:fvg_2D}) to the unobserved galaxy vector $\bd{g}_{ij}=(g_i, g_j)$ (brown vector in Fig. \ref{fig:fvg_2D}), as implied by Equation~\eqref{eq:flux_equals_ang_plus_galaxy}.
If we knew the galaxy vector $\bd{g}_{ij}$, we could clearly subtract it from the observed $(y_i(t),y_j(t))$ and obtain the desired AGN activity $(v_i(t), v_j(t))$. The unobserved vector $\bd{g}_{ij}$ defines a line $\bd{g}_{ij}\cdot x$, with $x\in\mathbb{R}$, that  intersects the dashed line in Figure~\ref{fig:fvg_2D} at point $\bd{x}_0\in\mathrm{R}^2$, which stands for the (not necessarily observed) minimum AGN activity at filters $i$ and $j$. Hence, if we can find $\bd{x}_0$, then we  know $\bd{g}_{ij}$.
The FVG method suggests the following: 
if we knew a vector $\bd{u}$ with the same direction as $\bd{g}_{ij}$, this would define the line $\bd{u}\cdot x,$ which is just a reparameterization of the line $\bd{g}_{ij}\cdot x$. Hence, the intersection of the line  $\bd{u}\cdot x$ 
with the dashed line would also occur at $\bd{x}_0,$ which would give us $\bd{g}_{ij}$. In a nutshell, FVG proposes that if we know the ratio of colors characterizing the host galaxy in question, then we can find the point $\bd{x}_0$ and separate the AGN and host galaxy luminosities.

In practical implementations of the FVG, the lines describing the total fluxes are identified using linear regression analysis. As pointed out in \cite{2012A&A...545A..84P}, the choice of the regression method is not trivial as the methods will fare differently depending on how well separated the dependent and independent variables are and depending on the intrinsic dispersion of data around the best fit and others factors (e.g., \citealt{1990ApJ...364..104I}; \citealt{1992MNRAS.257..659W}). The regression algorithms are based on the formulae provided by \cite{1990ApJ...364..104I}. The authors recommend a symmetric treatment of the variables using the ordinary least-squares (OLS) bisector method. A mean host-galaxy estimate is obtained from a random uniform distribution of possible host-galaxy values enclosed by the area that is formed by the intersection between a well-defined range of host-galaxy colors\footnote{PRM studies typically consider the host-galaxy slopes given by \cite{2010ApJ...711..461S} obtained for 11 Seyfert 1 galaxies.} and the AGN slopes derived from the bisector analysis. 

\subsection{Issues with previous FVG implementations}

\paragraph{Aggregating filters:}
As illustrated in Figure~\ref{fig:fvg_2D}, the FVG works on a 
flux-flux plot, defined by a pair of filters, and finds an 
intersection point $\bd{x}_0\in\mathrm{R}^2$. This is repeated for 
each pair of filters, and the resulting intersection points are then aggregated in some fashion or considered independently. In previous applications of the FVG, aggregation of the intersection points per filter pair has been done by averaging estimates  (e.g., \citealt{2012A&A...545A..84P}; \citealt{2015A&A...581A..93R}; \citealt{2019NatAs...3..251C}). Here we avoid this heuristic aggregation by considering  all filters simultaneously to infer $\bd{x}_0$. Hence, instead of working in a flux-flux space that contains information about two filters only, we  work  in the $F$-dimensional space of all filters (i.e., each filter has an axis) populated by tuples $(y_1(t), \dots, y_F(t))$,  and infer an intersection point $\bd{x}_0\in\mathrm{R}^F$.

\paragraph{Data uncertainty:}

The linear regression algorithms that were used in the past, such as the OLS bisector, assumed that the intrinsic dispersion of the data is larger than individual error measurements, hence the intersection point between the AGN and galaxy slopes, that is, the host galaxy, depends solely on the error range of the bisector fit. In this work, we account for the presence of noise in the data by explicitly taking error measurements into account in our model formulation (see Section~\ref{proposedmethod}). 

\paragraph{Uncertainty in the galaxy estimation:}

Previous FVG implementations provided point estimates for the intersection points and hence the galaxy contribution. By formulating a probabilistic version of FVG in this work, we are able to produce not just a point estimate for the intersection point, but rather a density of possible intersection points (see Section~\ref{intersect}). This density expresses our inability to pinpoint a unique intersection point due to the presence of noise (i.e., uncertainty) in the observed data, and acknowledges that a range of solutions are in fact plausible.

\subsection{Proposed probabilistic FVG}
\label{proposedmethod}

In this section we put forward a probabilistic reformulation of FVG based on probabilistic principal component analysis (PPCA; \citealt{tipping1999probabilistic}), which we call PFVG. In a first step, our method identifies the line formed by the total fluxes, and in a second step, it seeks the intersection of the unobserved galaxy line with the identified line.

We modeled the observed flux $y_i(t)$ in each filter $i\in \{1,\dots,F\}$ at time $t$ as a noisy observation from a latent signal $f(t)$\footnote{In the context of this work, the latent signal $f(t)$ correspond to the driving AGN continuum light curves modeled as a  random walk in Section~\ref{sec3}.},  common to all filters, scaled and shifted by filter-dependent parameters $a_i$ and $b_i$,
\begin{eqnarray}
y_i(t) = a_{i} f(t) + b_{i} + \epsilon_{i}(t) \ , \ i \in \{1,\dots,F\} \ , 
\label{eq:noisy_line}
\end{eqnarray}
where $\epsilon_{i}(t)$ is observational noise assumed to be drawn from a Gaussian distribution, $\epsilon_{i}(t)\sim \mathcal{N}(0,\sigma^2(t))$.
This incorporates the FVG assumption that a flux observation in a filter at time $t$ is linearly related to the observations in all other filters at time $t$.
Assuming observations co-occur, that is, they are observed at the same time instance, we  form vectors $\bd{y}(t) = (y_1(t), y_2(t),\dots, y_F(t))\in\mathbb{R}^F$ for each observed  $t$. Similarly, we also form $\bd{\sigma}^2(t) = (\sigma_1(t), \sigma_2(t),\dots, \sigma_F(t))$.
\subsubsection{Line identification}
The co-occurring, noisy observed total fluxes in Eq. \ref{eq:noisy_line} gives rise to the joint likelihood\footnote{\bd{I} stands for the identity matrix. Its dimensions are implicitly defined by its context, e.g., in Eq. \ref{eq:joint_likelihood}, it is of dimensions $F\times F$.},
\begin{eqnarray}
p(D|\bd{a}, \bd{b}, \bd{f}) = \prod_t \mathcal{N}(\bd{y}(t) | \bd{a} f(t) + \bd{b}, \bd{\sigma}^2(t)\bd{I})
\ ,
\label{eq:joint_likelihood}
\end{eqnarray}
where $\bd{a}=(a_1,\dots,a_F), \bd{b}=(b,\dots,b_F),$ and $D$ stands for all observed measurements, that is,~all available data. The proposed model seeks to find the line that goes through the observed total fluxes. It is essentially identical to PPCA in the special case where only one principal component is sought. Because we are not interested in the latent signal values $f(t)$, we treat them as nuisance variables: following PPCA, we impose on them a Gaussian prior $\mathcal{N}(\bd{f} |\bd{0},\bd{I})=\prod_{t} \mathcal{N}(\bd{f}(t) |0,1)$ and integrate them out, 
\begin{eqnarray}
p(D|\bd{a}, \bd{b}) &=& \int \prod_t \mathcal{N}(\bd{y}(t) | \bd{a} f(t) + \bd{b}, \bd{\sigma}^2(t)\bd{I})
\mathcal{N}(\bd{f} |\bd{0},\bd{I}) \bd{df}
\notag \\ 
&=& \prod_t \mathcal{N}(\bd{y}(t) | \bd{b}, \bd{a}\bd{a}^T + \bd{\sigma}^2(t)\bd{I}) \ .
\label{eq:log_likelihood_marginalised_f}
\end{eqnarray}
Maximizing $p(D|\bd{a}, \bd{b})$ in Equation \ref{eq:log_likelihood_marginalised_f} with respect to line parameters \bd{a} and \bd{b} gives us point estimates for these parameters. However, we wish to obtain a posterior density for these parameters because we are interested in the uncertainty of our estimates. Following Bayesian principal component analysis (\citealt{bishop1999}; \citealt{export:67241_rd}), we treat \bd{a} and \bd{b} as random variables and impose priors $\mathcal{N}(\bd{a}|\bd{0}, \nu_\alpha\bd{I})$
and $\mathcal{N}(\bd{b}|\bd{0}, \nu_b\bd{I})$\footnote{In all numerical experiments, we use $\nu_\alpha = \nu_\beta = 10^{-4}$.}. We now seek to obtain the posterior density $p(\bd{a}, \bd{b}|D) \propto p(D|\bd{a}, \bd{b}) p(\bd{a}) p(\bd{b})$.

Unfortunately, the posterior density $p(\bd{a}, \bd{b}|D)$ cannot be calculated analytically.
Instead, we seek a Gaussian approximating posterior $\mathcal{N}(\bd{a}, \bd{b}|\bd{\mu}, \bd{\Sigma})$ that is as close as possible to the exact posterior $p(\bd{a}, \bd{b}|D)$ (\citealt{export:67241_rd}). An appropriate notion of closeness is given by the Kullback-Leibler divergence between the two densities, $D_{KL}(\mathcal{N}(\bd{a}, \bd{b}|\bd{\mu}, \bd{\Sigma})||p(\bd{a}, \bd{b}|D))$. 
By minimizing this divergence with respect to the free parameters $\bd{\mu}$ and $\bd{\Sigma}$, we obtain the optimal (i.e.,~closest) Gaussian approximation to $p(\bd{a}, \bd{b}|D)$.

\subsubsection{Intersection}
\label{intersect}
The Gaussian approximating posterior $\mathcal{N}(\bd{a}, \bd{b}|\bd{\mu}, \bd{\Sigma})$ presents not just a single solution for the parameters of the line formed by the total fluxes, but rather a density of possible parameter solutions and hence a density of possible lines. Our goal here is to work out the density of intersection $\bd{x}_0$ between a given candidate line $\bd{u}\cdot x$ for the unobserved galaxy and the density of lines implied by $\mathcal{N}(\bd{a}, \bd{b}|\bd{\mu}, \bd{\Sigma})$.

While the term intersection has a clear meaning for two lines that live in a Euclidean space,  we still need to clarify what we mean by it in our setting.
We consider a point $\bd{x}_0$ to be an intersection if it simultaneously satisfies two conditions, namely: (1) it is close to the line $\bd{u}\cdot x$, and (2) it is close to a line $\bd{a}\cdot t + \bd{b,}$ which enjoys high density according to $\mathcal{N}(\bd{a}, \bd{b}|\bd{\mu}, \bd{\Sigma})$. Allowing momentarily for some noise tolerance $\varsigma$ and $\varrho$, we express the joint satisfaction of these two conditions as the product $\mathcal{N}(\bd{x}_0|\bd{u}\cdot x, \varsigma^2) \cdot \mathcal{N}(\bd{x}_0|\bd{a}\cdot t + \bd{b}, \varrho^2)$
where each term stands for conditions (1) and (2), respectively.
Noise $\varsigma$ and $\varrho$ control how close $\bd{x}_0$ must be to the two lines in order to be considered an intersection. Hence, we  view the two terms  as soft constraints that need to be satisfied.
By driving $\varsigma$ to $0$, the first constraint becomes a hard constraint that states that $\bd{x}_0$ must be on the line $\bd{u}\cdot x$. Roughly speaking, the product now becomes $\lim_{\varsigma \to 0}
\mathcal{N}(\bd{x}_0|\bd{u}\cdot x, \varsigma^2\bd{I}) \mathcal{N}(\bd{x}_0|\bd{a}\cdot t + \bd{b}, \varrho^2\bd{I}) = \mathcal{N}(\bd{u}\cdot x |\bd{a}\cdot t + \bd{b}, \varrho^2\bd{I})$.

\begin{figure}
 \includegraphics[width=\columnwidth]{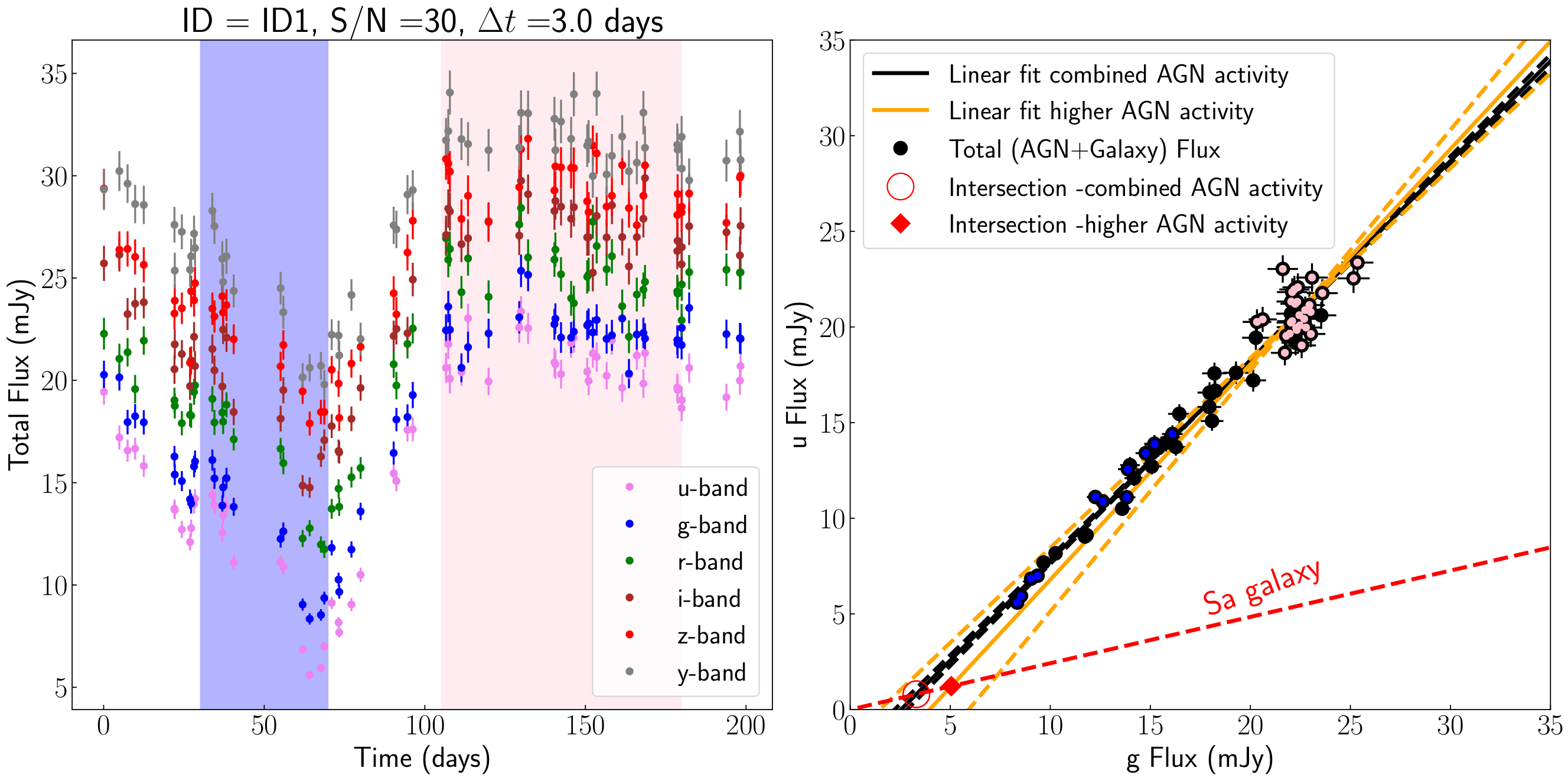}
 \caption{Flux variation gradient applied to set of light curves from our LSST mock catalog. The left panel shows the light curves with an average time sampling of $\Delta t = 3$ days, $\rm{S/N}=30$, and total time span of $T = 200$ days. Low and high levels of AGN activity are marked with blue and magenta boxes, respectively. The right panel shows the FVG analysis for the $u$ and $g$ bands only. Other bands are given in the appendix in Figure~\ref{fig:Apx1}. The dotted orange and black lines cover the upper and lower standard deviations of the OLS bisector fit (solid lines) for the combined (black) and higher (orange) AGN state of activity. The color of the Sa galaxy we used in the simulations is shown as a dotted red line.}
 \label{fig:fvgexa}
\end{figure}

Hence, given $\bd{a}$, $\bd{b,}$ and  $t$, the conditional support for $\bd{u}\cdot x$ being an intersection point, or equivalently, the conditional support for  coordinate\footnote{Coordinate $x$ addresses a unique point on the line $\bd{u}\cdot x$.}  $x$ addressing an intersection point is $p(x|\bd{a},\bd{b},t) = \mathcal{N}(\bd{u}\cdot x |\bd{a}\cdot t + \bd{b}, \varrho^2\bd{I})$.
Assuming the improper prior\footnote{Our method does not rely on this particular choice of prior.} $p(t)\propto 1$, the unconditional support for $x$ is
\begin{align}
&p(x) \propto \int\int\int p(x|\bd{a},\bd{b},t) \mathcal{N}(\bd{a},\bd{b}|\bd{\mu}, \bd{\Sigma}) \ p(t) \ \bd{da}\bd{db} dt  \notag \\
& = \int\int\int \mathcal{N}(\bd{u}\cdot x | \bd{a}\cdot t + \bd{b}, \varrho^2\bd{I})   \mathcal{N}(\bd{a},\bd{b}|\bd{\mu}, \bd{\Sigma})  p(t) \  \bd{da}  \bd{db} dt \ .
\label{eq:intersection_integral}
\end{align}
 Specifying the following partitions of mean and covariance,
 \begin{align}
 \bd{\mu}=\begin{pmatrix} \bd{\mu}_a \\ \bd{\mu}_b \end{pmatrix} \ ,  \
 \bd{\Sigma} = \begin{pmatrix} \bd{\Sigma}_{aa} && \bd{\Sigma}_{ab} \\ \bd{\Sigma}_{ba} && \bd{\Sigma}_{bb} \end{pmatrix} \ ,
 \end{align}
 as well as $\bd{\Sigma}_{a|b}= \bd{\Sigma}_{aa} - \bd{\Sigma}_{ab}\bd{\Sigma}^{-1}_{bb}\bd{\Sigma}_{ba}$,  integration  over $\bd{a}$ and $\bd{b}$ yields
\begin{align}
& p(x) \propto \notag\\
&\int \mathcal{N}(\bd{u}\cdot x | \bd{\mu}_a\cdot t + \bd{\mu}_b, \varrho^2\bd{I} + \bd{\Sigma}_{a|b}t^2 +\bd{R}(t) \bd{\Sigma}_{bb} \bd{R}(t)^T) p(t) dt  \ ,
\label{eq:intersection_second_integral}
\end{align}
where we have defined $\bd{R}(t)=\bd{\Sigma}_{ab}\bd{\Sigma}^{-1}_{bb}t+\bd{I}$. Again, by driving noise $\varrho$ to 0, we finally arrive at the support for $x$ given by the density of lines implied by $\mathcal{N}(\bd{a},\bd{b}|\bd{\mu}, \bd{\Sigma})$ in the absence of noise (i.e.,~ $\varsigma=\varrho=0$)\footnote{In practice, we  set them to a low value, e.g.,  $\varsigma=\varrho=10^{-10}$ for numerical stability.}:
\begin{align}
p(x) \propto 
\int \mathcal{N}(\bd{u}\cdot x | \bd{\mu}_a\cdot t + \bd{\mu}_b, \bd{\Sigma}_{a|b}t^2 +\bd{R}(t) \bd{\Sigma}_{bb} \bd{R}(t)^T) p(t) dt  \ .
\label{eq:intersection_third_integral}
\end{align}
Unfortunately, we cannot  calculate the integral in Equation \ref{eq:intersection_third_integral} analytically.
To make progress, we proceed in two steps: first, we approximate the integrand with the Gaussian approximation $\mathcal{N}(x,t|\bd{m},\bd{S})$ by minimizing the Kullback-Leibler divergence between them. Second, we calculate the Gaussian marginal $\mathcal{N}(x|m_{x},s^2_{x})=\int\mathcal{N}(x,t|\bd{m},\bd{S}) dt,$
which is the approximate support for coordinate $x$. 

Hence, the density that describes the distribution of the $i-$th coordinate of the sought intersection $\bd{u}x$ is the Gaussian $\mathcal{N}(u_i x|u_i m_{x},u_i^2 s^2_{x})$.
We note, however, that the Gaussian density allows negative values for coordinate $x$. To avoid this, we  work  with the reparameterization $\chi=\sqrt{x}$, that is,~we infer an approximate Gaussian marginal posterior $\mathcal{N}(\chi|m_{\chi},s^2_{\chi})$ for $\chi$ instead of $x$. 
Consequently, because $\chi$ has a Gaussian distribution, coordinate $x$ has a noncentral chi-squared distribution that supports only positive values (\citealt{johnson1995continuous}). 

Finally, because of the complex probability density of the noncentral chi-squared distribution, we define the density of the $i-$th coordinate of the sought intersection $\bd{u}x$ implicitly as

\begin{equation}
\label{eq:q_i}
u_i x = u_i \chi^2, \chi\sim \mathcal{N}(\chi|m_{\chi},s^2_{\chi}) \ ,\end{equation}

\begin{figure}
 \centering
  \includegraphics[width=\columnwidth]{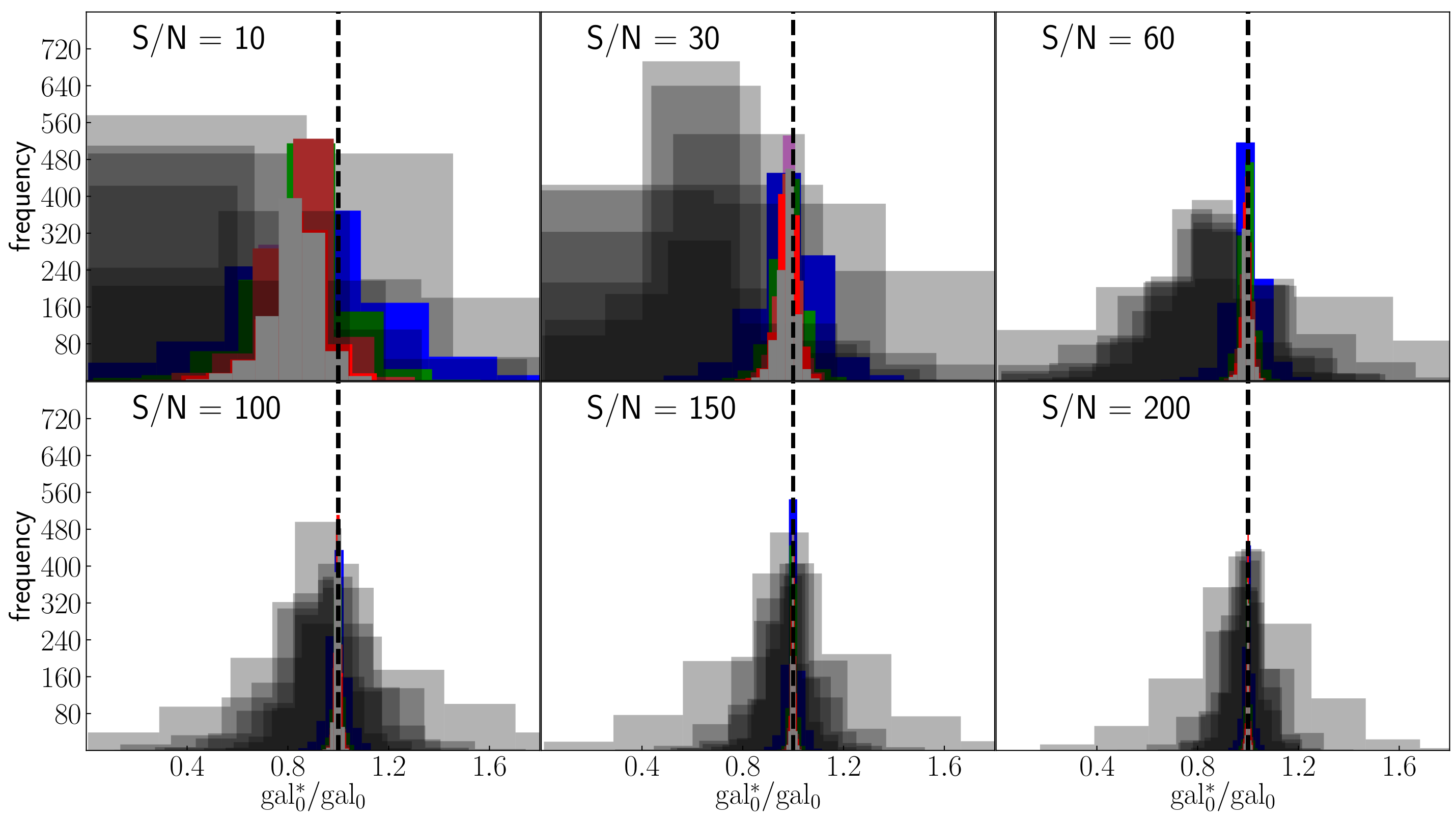}
 \caption{Recovered FVG host-galaxy fluxes ($\rm gal^{*}_{0}$) for various S/N and time sampling. The distributions obtained for the CS are shown with the same colors as the light curves in Figure~\ref{fig:fvgexa}. To facilitate comparison, we show in black the distributions obtained for the HS. The FVG applied to the CS appears to be considerably better than the HS; it is characterized by the smallest dispersion around the true galaxy ($\rm gal_{0}$; vertical dotted black line) for an S/N > 100 (see text).
 }
 \label{fig:fvghisto}
\end{figure}

\begin{figure*}
\begin{tabular}{cc}
  \includegraphics[width=85mm]{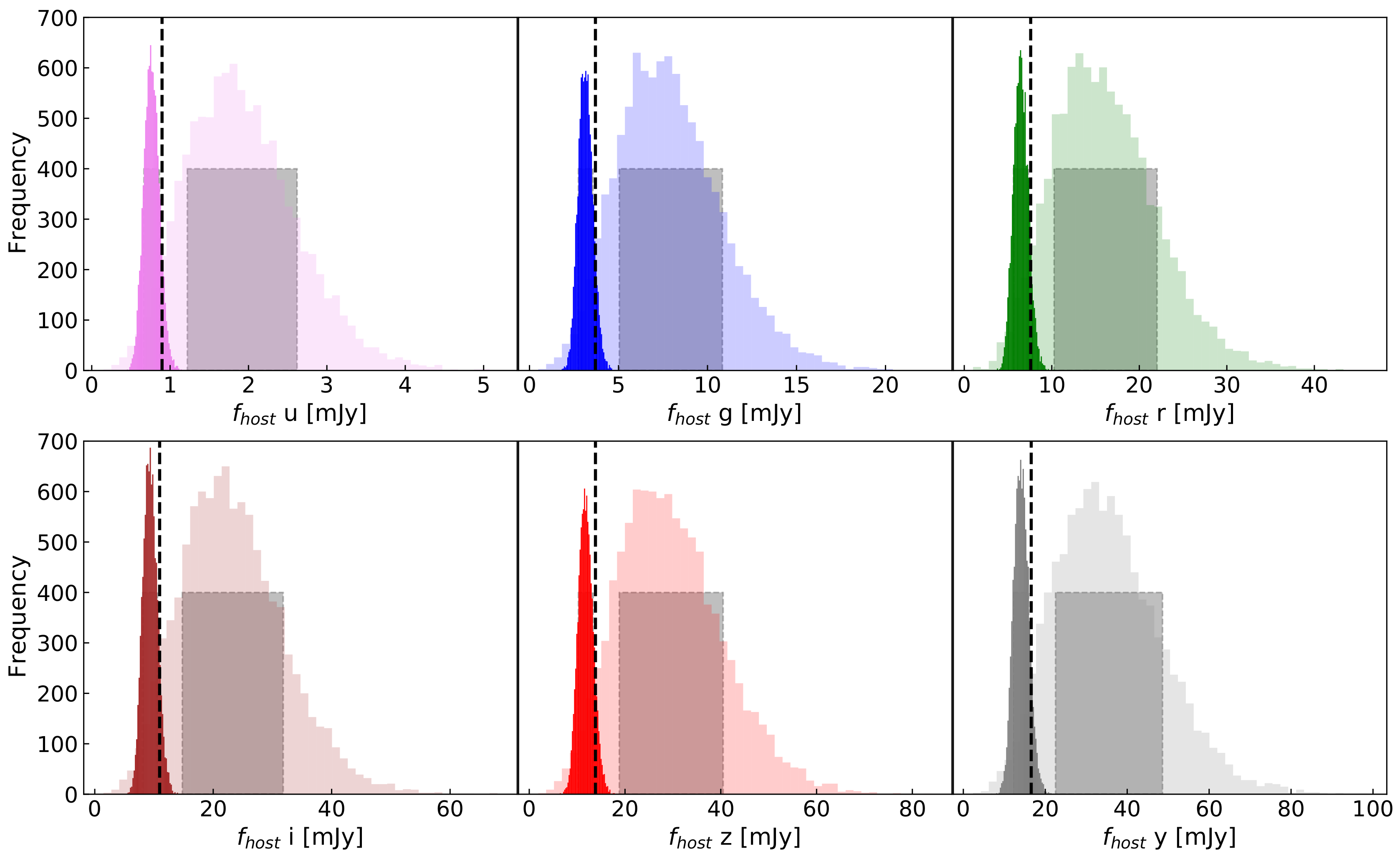} &   \includegraphics[width=85mm]{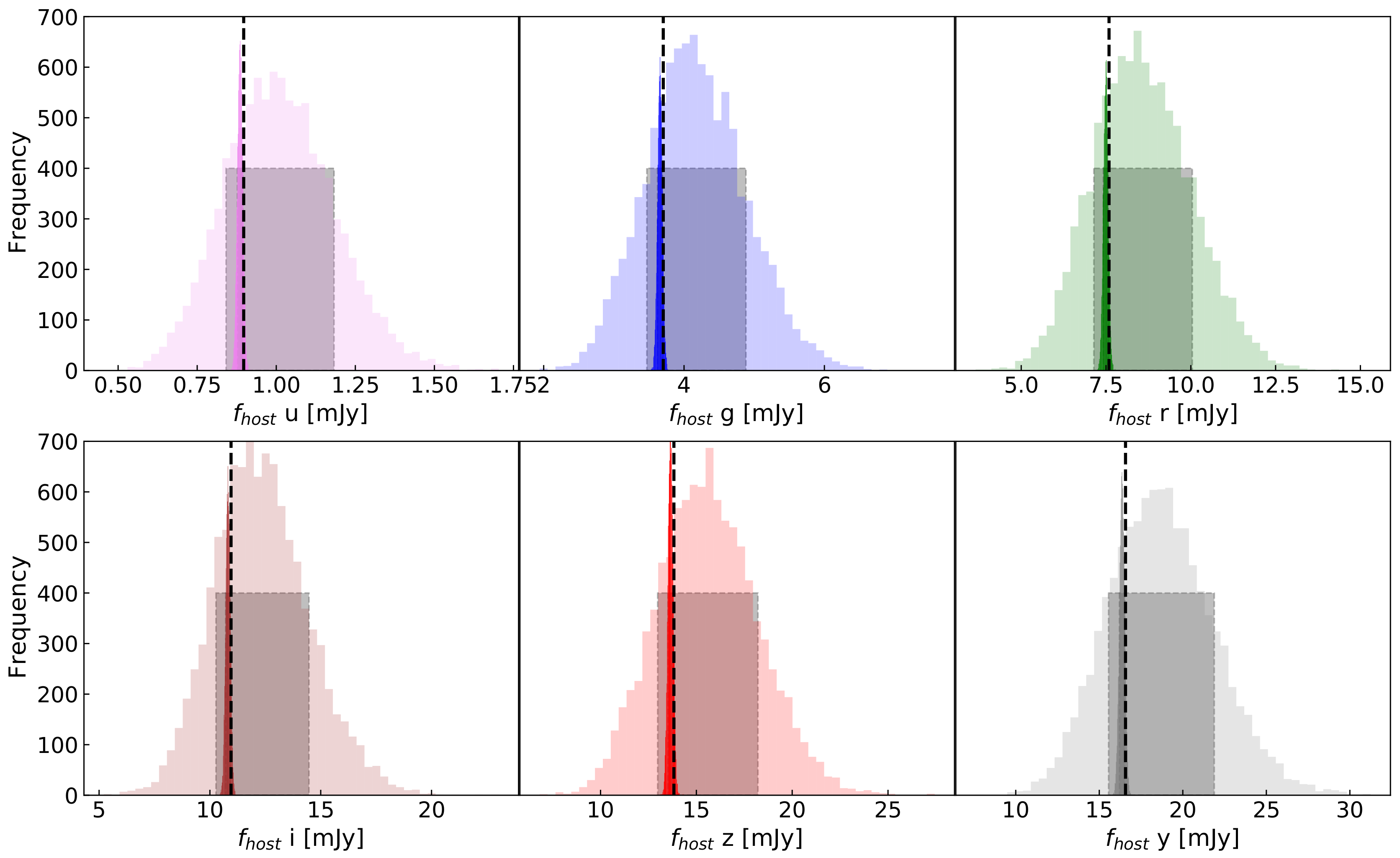} \\
(a) S/N = 10, $\Delta t= 10$ days  & (b) S/N = 100, $\Delta t= 10$ days \\[6pt]
 \includegraphics[width=85mm]{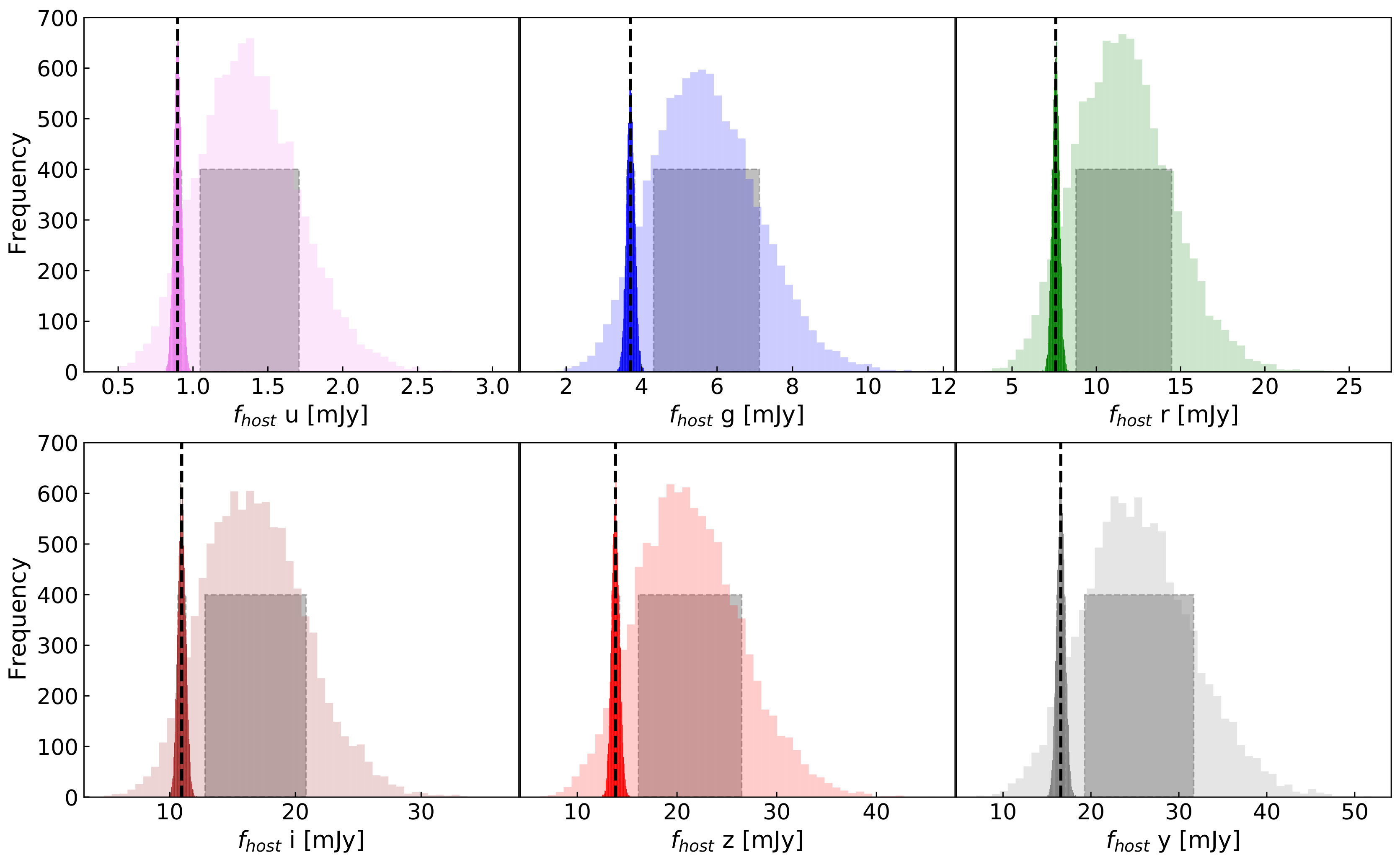} &   \includegraphics[width=85mm]{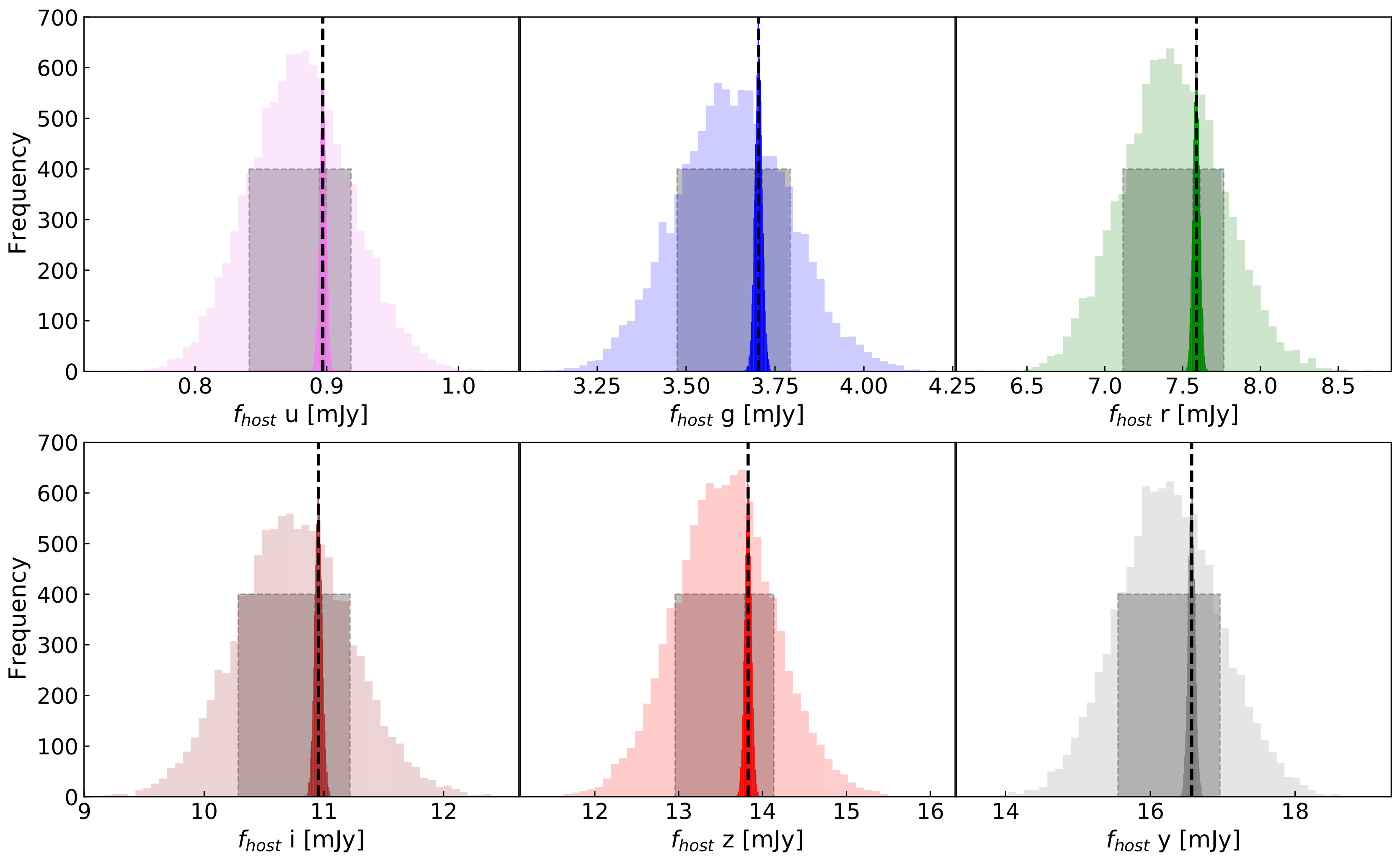} \\
(c) S/N = 10, $\Delta t= 1$ days & (d) S/N = 100, $\Delta t= 1$ days  \\[6pt]
\end{tabular}
\caption{Recovered PFVG distributions (see Equation \ref{eq:q_i}) of host-galaxy fluxes obtained for the HS (transparent colored histogram) and CS (solid colored histogram). Distributions obtained for poorly sampled light curves, with low and high S/N, are shown in panels (a) and (b), respectively. Distributions obtained for well-sampled light curves, with low and high S/N, are shown in panels (c) and (d), respectively.  The vertical dotted line marks the true galaxy value used in the simulations ($\rm gal_{0}$). The 68\% confidence range used to estimate the 1$\sigma$ uncertainty around the median is shown as transparent black boxes.}
\label{fig:PFVGselect}
\end{figure*}

\section{Simulations}\label{sec3}

In this section we explore the ability of the FVG, along with our probabilistic reformulation, to recover the host-galaxy contribution under different simulated scenarios. We consider the observational characteristics of PRM campaigns such as time sampling, signal-to-noise ratio (S/N), and the occurrence of time delays in the observed light curves. In particular, we focus on the next future public optical or near-infrared Legacy Survey of Space and Time (LSST), which will monitor about ten million quasars in six broad-band filters ($ugrizy$) during ten years of operation.

Our simulation method follows that of \cite{2019MNRAS.490.3936P}. First, we modeled the driving AGN continuum light curves as a random walk process with a power spectral density $P(\nu)\propto \nu^{-2}$ (\citealt{2009ApJ...698..895K}; \citealt{2017ApJ...834..111C}). The total emission $L(\rm AGN)_{\nu}$ from the AGN was then calculated by integrating the Planck function of a black body, 

\begin{align}
L({\rm AGN})_\nu = 2 \pi \int_{r_\mathrm{in}}^{r_\mathrm{out}} B_\nu(T(r)) r\mathrm{d}r,
\label{equ:AGNspec}
\end{align}
where $B_\nu$ is characterized by the radial temperature profile of an optically thick geometrically thin accretion disk, $T(r)\propto r^{-3/4}$. The effects of varying the disk luminosity as a function of orientation are negligible for the purpose of this work. Consequently, we assume in Equation~\ref{equ:AGNspec} a face-on ($i=0^{\circ}$) configuration.
We then mixed the AGN accretion disk emission with the host-galaxy contribution measured from the UV/optical galaxy templates obtained by \cite{1996ApJ...467...38K} for different morphology types. Finally, the total observed fluxes were calculated by convolving each model component with the transmission curves of the LSST broad-band filters.
In Figure~\ref{fig:adgalspec} we show an example of the random walk simulated light curves, AGN accretion disk, and host-galaxy spectrum for an arbitrary local Seyfert galaxy from our mock catalog.

\subsection{Effects of time sampling, S/N, and AGN activity}
\label{sec:III}

The light curves of AGN have a stochastic nature, and it is very likely to encounter cases in which either a low or a high level of activity is observed. Combined with the fact that astronomical observations are frequently affected by weather conditions, technical issues, and seasonal gaps, this poses significant challenges to the FVG method. In Figure~\ref{fig:fvgexa} we present an example that illustrates the impact of the above issues on the traditional FVG analysis. The time sampling, $\Delta t$, during the LSST monitoring is expected to be between 2 to 5 days. Thus, the light curves were randomly sampled with an average sampling of $\Delta t = 3$ days. In this example, we adopted an  $\rm{S/N}=30$, which corresponds to the highest expected photometric noise, in particular, for the $u$ band (at $\sim3\%$  noise level). The left panel shows the light curves depicting two distinct states of AGN activity. A low state characterized by small variability amplitudes, with fluxes undergoing a steep decrease of about 40\%\  before reaching a minimum at day 65. Afterward, the flux starts to increase by about the same amplitude (40\%) until a maximum is reached at 100 days. This represents the beginning of the higher state of AGN activity. At this point, the luminosity has already increased by a magnitude of about 1.5 in all filters. During this state, the AGN exhibit a period of stronger variability with light curve features characterized by single outbursts with $\sim10\%$  variability amplitudes, as seen in the range between 140 to 180 days. These variability events, regulated by both low and high states of AGN activity, are crucial for the performance of the FVG method.
On the one hand, identifying a line requires that we know any two points. However, because the points are noisy, not every pair of points leads to a reliable identification of the line. If we had the ability of measuring both the highest activity flux and the lowest activity flux, we would obtain the two most distant points in the multidimensional flux space. This particular pair of points would lead to a reliable identification of the line we are looking for. This can be understood intuitively; perturbing one of these two points by noise would only perturb the line slightly. In other words, the more of the dynamic flux range we observe, the more reliably we can estimate the line.
In contrast, if we only measure two fluxes of average activity, this would give us two points in the multidimensional flux space that are very close together. This pair of points would lead to a nonrobust identification of the sought line; perturbing one of these points by noise would perturb the line significantly. In other words, the less of the dynamic flux range we observe, the less robustly we can estimate the line. To quantify the impact of the observed dynamic range in flux, we applied the FVG method on low and high states of AGN activity. These states are shown in the FVG diagram in the right panel of Figure~\ref{fig:fvgexa}.

In a first test, we simulated the aforementioned situation where we observe a large part of the dynamic flux range. Therefore we considered the combined AGN states (hereafter denoted as CS) and performed an OLS bisector linear regression analysis. The bisector method yields a linear gradient of $\Gamma_{ug} = 1.04\pm0.02$, which is consistent with $\Gamma_{UB} \sim1$ derived from observations of Seyfert 1 galaxies (e.g., \citealt{1992MNRAS.257..659W}; \citealt{2010ApJ...711..461S}). The intersection point (red circle in Figure~\ref{fig:fvgexa}) between the AGN and galaxy slopes gives a host-galaxy flux of $f^{\rm host}_{u} = 0.80$ mJy and $f^{\rm host}_{g} = 3.31$ mJy. Here the true host-galaxy vector, $\rm gal_{0} = (0.89,3.70)$\footnote{We denote the true galaxy vector by $\rm gal_{0}$, which is represented by the FVG intersection point of the noiseless and ideally sampled light curve ($\Delta t = 0.01$ days), as shown in the appendix in Figure~\ref{fig:Apx1}. The photometric bands implied by $\rm gal_{0}$ are made clear by the context.}, obtained from the scaled Sa-type template (Figure~\ref{fig:adgalspec}), is recovered at $10$ and $10.5\%$  precision for $u$ and $g$ bands, respectively. 

In a second test, we considered only the higher state of AGN activity (hereafter denoted as HS). In this case, we simulated the situation in which we observe a smaller part of the dynamic flux range. The bisector method yields a linear gradient of $\Gamma_{ug} = 1.13\pm0.13$ (orange line). The AGN slope is notably steeper, and with a larger uncertainty, hence leading to a different intersection point ($f^{\rm host}_{u} = 1.22$ mJy and $f^{\rm host}_{g} = 5.05$ mJy; red diamond), and overestimating the true host galaxy by about $\sim40\%$  in both $u$ and $g$ bands.

\begin{figure}
    \centering
    \includegraphics[width=\columnwidth]{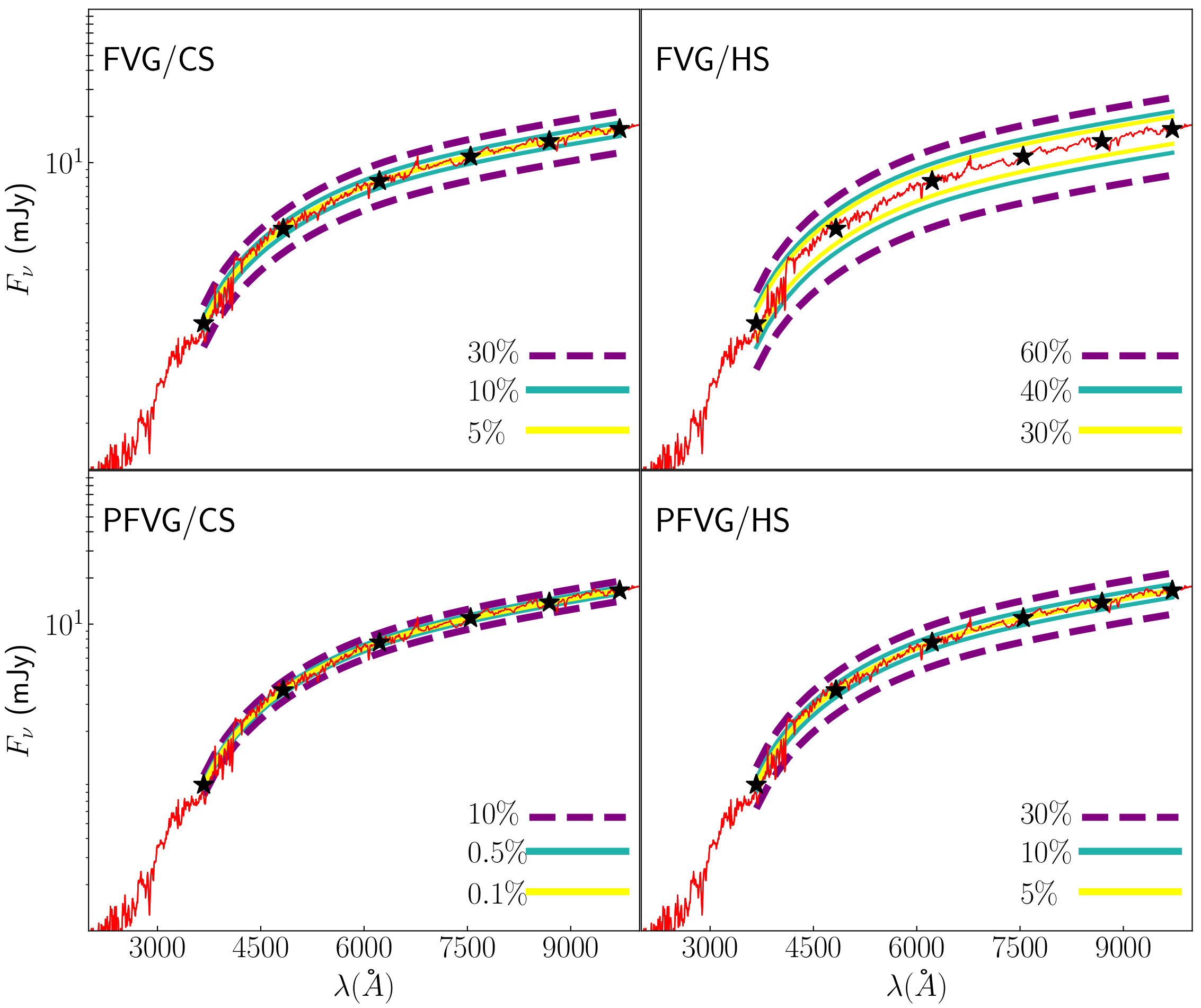}
    \caption{Precision on the recovered galaxy fluxes obtained from the FVG (upper panels) and PFVG (lower panels) distributions for the CS and HS of AGN activity. The red line shows the Sa-type host-galaxy (observer's frame) spectrum. The galaxy fluxes (in mJy) at each filter are plotted as filled black stars. The lines indicate the range of galaxy values that is recovered within the 68\% confidence interval, and they are given for an S/N = 10 (purple), 100 (cyan), and 200 (yellow).}
    \label{fig:SED_FVGmock}
\end{figure}

To study the performance of the FVG under different observational conditions, we repeated the same analysis for simulated light curves with various S/Ns (10, 30, 60, 100, 150, and 200) and time sampling chosen in the interval $[\Delta t_{\rm min},\Delta t_{\rm max}$] = $[0.1,10]$ spaced by 0.01 days. We denote by $\rm gal_{0}^{*}$ the estimated host-galaxy vector\footnote{As in the definition of $\rm gal_{0}$, the implied photometric bands are made clear by the context.}. The recovered distributions of host-galaxy fluxes, $\rm gal_{0}^{*}$, are shown in Figure~\ref{fig:fvghisto}. We find that the performance of the FVG when applied to the HS is strongly dependent on the quality of the light curves; in particular, the FVG appears more sensitive to the presence of noise rather than to time sampling. In the case of S/N < 100, the HS (black histograms) exhibits a broad and asymmetric distribution around the central values, with a systematic bias (up to about 50\%\ ) toward small $\rm gal^{*}_{0}$. For higher-quality data (i.e., S/N $\geq 100$), the distributions are more symmetric and unbiased with $\langle \rm gal^{*}_{0} \rangle \simeq \rm gal_{0}$, but they are still substantially broader around the mean value. In contrast to the HS, the FVG performance on the CS (colored histograms) becomes less sensitive to the S/N, leading to improved galaxy statistics with a relatively unbiased and narrower distributions of $\rm gal_{0}^{*}$. An exception is seen for the lower quality data (i.e., S/N $\leq 30$), where a bias (up to about 30\%) toward lower values is still present. For the case of S/N $\geq 100$, the CS distributions becomes narrower and the true galaxy is recovered with a precision of 10\%. Based on these simulations, the recovered precision is consistent with the precision obtained from PRM observational campaigns. We return to this issue in Section~\ref{sec:data}.

\begin{figure}
    \centering
    \includegraphics[width=\columnwidth]{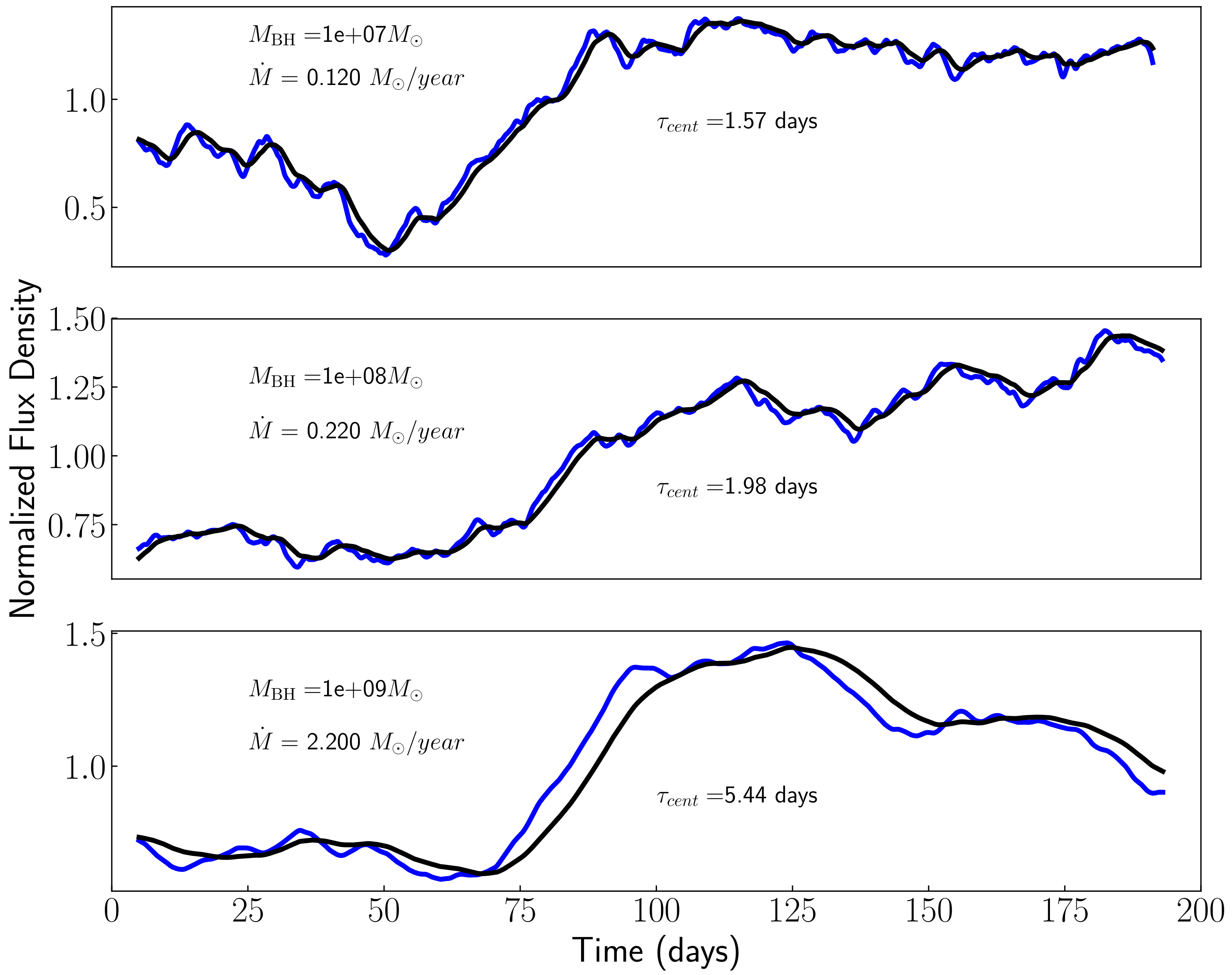}
    \caption{AD mock light curves with time delays. For illustration, only the $u$ (blue) and $y$ (black) light curves are shown. The former have the longer delay. The centroid of the transfer function obtained for the y band with respect to the u band is labeled inside each panel.}
    \label{fig:lcfvgdelay}
\end{figure}

\subsection{Comparison with the proposed PFVG method}
\label{sec:pfvgcomp}

We have carried out the PFVG analysis using the same simulated light curves as for the traditional FVG in the previous section. The results are summarized in Figure~\ref{fig:PFVGselect}. The PFVG performance on the CS appears to be considerably superior to the traditional FVG, especially for the case of the worst quality data (case a in Figure~\ref{fig:PFVGselect}; S/N=10, $\Delta t=10$ days); it exhibits a very narrow distribution around the central value, with the true galaxy underestimated by only 15\%\  compared to the HS. In this particular case, the HS deviates from the true galaxy by about twice the value. The PFVG performance substantially improves already at S/N = 100 (panels b and d) where the HS exhibits a more symmetric distribution with the true galaxy within the  $1 \sigma$ sigma range. Notably, the CS distributions are significantly narrower and the true galaxy is recovered with an exceptional precision of 0.5\%. We attribute the good performance of the PFVG to the fact that it accounts for the data measurement errors $\sigma_i(t)$ in its formulation (see Eq.~\ref{eq:joint_likelihood}). As a consequence, we see that as S/N decreases, the performance of PFVG deteriorates in a graceful manner rather than abruptly. A comparison between the recovered FVG and PFVG galaxy fluxes is shown in Figure~\ref{fig:SED_FVGmock}. The PFVG results are summarized in Table~\ref{pfvgval}.

\subsection{PRM light curves with time delays}

Time delays, $\tau_{c}(\lambda)$, between different UV/Optical continuum bands are a prediction of the AD thermal reprocessing scenario ($\tau_{c} \propto \lambda^{4/3}$; \citealt{1998ApJ...500..162C}; \citealt{2002apa..book.....F}; \citealt{2007MNRAS.380..669C} ). The delays are interpreted in terms of the light travel time across different regions of the AD, and have been detected for a few sources over the past years. The reported measurements are found to be overestimated by factor of $\sim3$ than the expected for an optically thick geometrically thin AD (see \citealt{2019MNRAS.490.3936P} and references therein). This overestimation has been attributed mainly to contamination of broad emission lines into the bandpasses, including additional diffuse continuum from the BLR (\citealt{2019NatAs...3..251C}; \citealt{2019MNRAS.489.5284K}). Regardless of their magnitude, there is growing evidence that supports their existence in most, if not all, AGN. Whether their production mechanism is exclusively due to AD continuum contribution is a matter of current debate.

\begin{figure}
    \centering
    \includegraphics[width=\columnwidth]{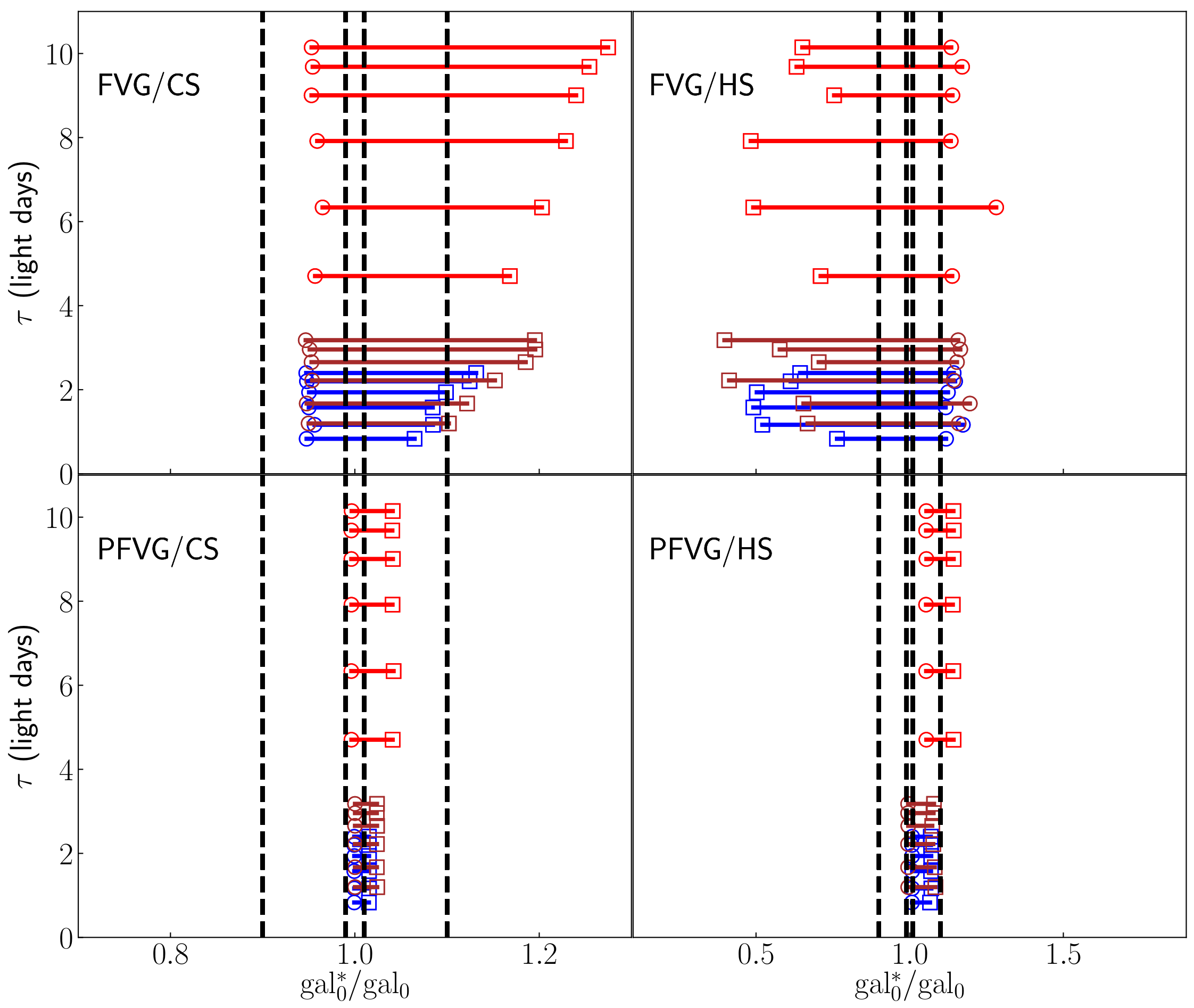}
    \caption{Distributions of host-galaxy fluxes ($\rm gal_{0}^{*}$) obtained by the FVG (top) and PFVG (bottom) as a function of time delays. The distributions are shown for each object (in blue, brown, and red) obtained for different S/N, filters, and for a fixed time-sampling $\Delta t = 3$ days. The open circles and squares mark an S/N = 100 and 30, respectively. Dashed inner and outer lines represent 1 and 10\% precision, respectively.}
    \label{fig:delayFVGPFVG}
\end{figure}

In order to quantify any potential biases that might be introduced by PRM light curves with time delays, we generated mock light curves of the continuum based on the AD thermal reprocessing scenario, as outlined in \cite{2019MNRAS.490.3936P}. In brief, the observed AD UV or optical continuum emission, $F_{c}(\lambda, t)$, is the result of the convolution of the X-ray driving light curve, $F_{x}(t)$, with a response function $\Psi(\tau|\lambda) \propto \partial B_{\nu}(\lambda,T(t-\tau)) / \partial L_{x}(t-\tau)$, so that $F_{c}(\lambda, t) = F_{x}(t) * \Psi(\tau|\lambda)$, where $\tau \propto (r^2+h^2)^{1/2}$ is the time-delay function for a face-on Keplerian ring or disk structure. The disk is irradiated by an X-ray corona located at a distance $h$ above the plane  (\citealt{2005ApJ...622..129S}) and has a temperature profile, $T(r)\propto (M_{\rm BH} \dot M)^{1/4}r^{-3/4}$, given by the standard \cite{1973A&A....24..337S} theory. 

We generated the AD mock light curves for three objects with black hole masses $M_{\rm BH} = 10^{7}$, $10^{8}$, and $10^{9} M_{\odot}$ and accretion rates $\dot M = 0.12$, $0.22$, and $2.2 \, M_{\odot} yr^{-1}$, respectively. We assumed that the objects are accreting at 10\% Eddington. We used the same random seed  as in Figure~\ref{fig:fvgexa} to create the driving X-ray light curve. The driver was then convolved with transfer functions obtained for each black hole mass and accretion rate. We note that the choice of these model parameters is only for illustrative purposes, allowing the time delay to be long enough to quantify possible biases while neglecting the effects of redshift and galaxy contribution dependence. 

Figure~\ref{fig:lcfvgdelay} shows an example of the simulated AD light curves observed in the the $u$ and $y$ bands. 
These two bands display the largest difference between them in terms of time delay, as measured by the centroid of the transfer functions.
The original light curves have an ideal time sampling of $\Delta t = 0.1$ days and a time span of 190 days. We added the galaxy contribution, observational noise, and resampled the light curves in the same way as performed in Section~\ref{sec:III}. The FVG and PFVG were applied on the resampled and noisy light curves. The recovered distributions of the host galaxy, $\rm gal_{0}^{*}$, as a function of $\tau$ for different bands are shown in Figure~\ref{fig:delayFVGPFVG}. The results are shown for a fixed time sampling of $\Delta t =3$ days and for an S/N = 30 and 100. The analysis includes both CS and HS cases of AGN activity. We find that the traditional FVG is more sensitive to the presence of time delays in the light curves; in the CS case, it shows a clear bias toward higher galaxy values by up to $\sim20$\% as the time delay increases and the S/N decreases (open squares). A more stable behavior is seen for S/N = 100 (open circles), where the galaxy is underestimated by only $\sim5$\% of the true value. The opposite behavior is seen for the HS case, in which the galaxy is biased toward lower values as the S/N decreases. In this case, the galaxy is underestimated by about 50\% for an S/N = 30 and overestimated by up to $\sim15$\% for an S/N = 100.

The bottom panel of Figure~\ref{fig:delayFVGPFVG} shows that the proposed PFVG method is less sensitive to time delays in the light curves, and it clearly outperforms the FVG in CS and HS cases of AGN activity. It shows considerably less biased results and a more stable behavior per filter. The CS case is most notable, where the host galaxy is recovered within a precision of $\sim$1 and 5\% for S/Ns of 100 and 30, respectively. In the HS case, the magnitude of the bias increases up to $\sim$10\% of the true value.

\section{Data applications}
\label{sec:data}
In this section we apply the PFVG method to AGN data obtained by PRM campaigns, for which host-galaxy estimates have been obtained with the traditional FVG method. We select three sources: Mrk509, Mrk279, and 3C120. The Mrk objects were monitored as part of a narrow-band PRM campaign dedicated to the RM of the accretion disk in AGN (\citealt{2017PASP..129i4101P}; \citealt{2019NatAs...3..251C}). The object 3C120 was observed as part of a simultaneous broadband PRM monitoring of the BLR and the dust torus in AGN (\citealt{2018A&A...620A.137R}). In the following cases, the candidate host-galaxy vector (denoted by \bd{u} in Sect~\ref{intersect}) is given by \cite{2010ApJ...711..461S} unless stated otherwise. All  data are publicly available and can be retrieved directly from the respective journals.

\begin{figure}
    \centering
    \includegraphics[width=\columnwidth]{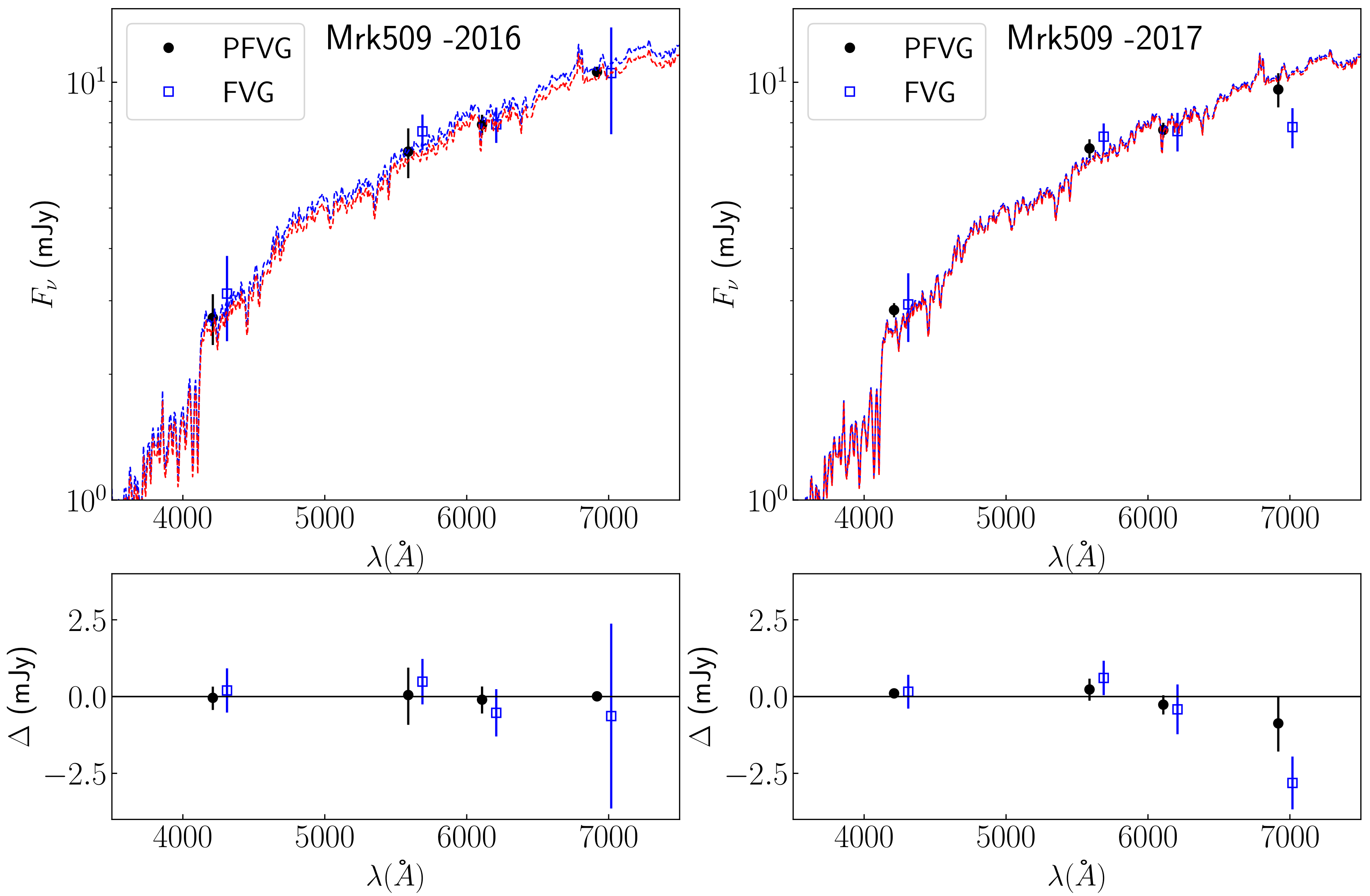}
    \caption{Application of the PFVG method to the object Mrk509 for epochs 2016 (left) and 2017 (right). The black dots correspond to the median of the recovered distributions of host galaxy fluxes (see Appendix~\ref{fig:Apx3}). The error bars show the formal 1$\sigma$ uncertainty around the median. The results obtained by the FVG are shown with open blue squares. The redshifted bulge galaxy template of \cite{1996ApJ...467...38K} is fitted to both PFVG (red) and FVG (blue) host-galaxy fluxes. The fit residuals, as obtained for the PFVG and FVG, are shown in the bottom panels. A small wavelength shift between the PFVG and FVG was introduced for clarity.}
    \label{fig:AppMrk509}
\end{figure}

\begin{figure}
    \centering
    \includegraphics[width=\columnwidth]{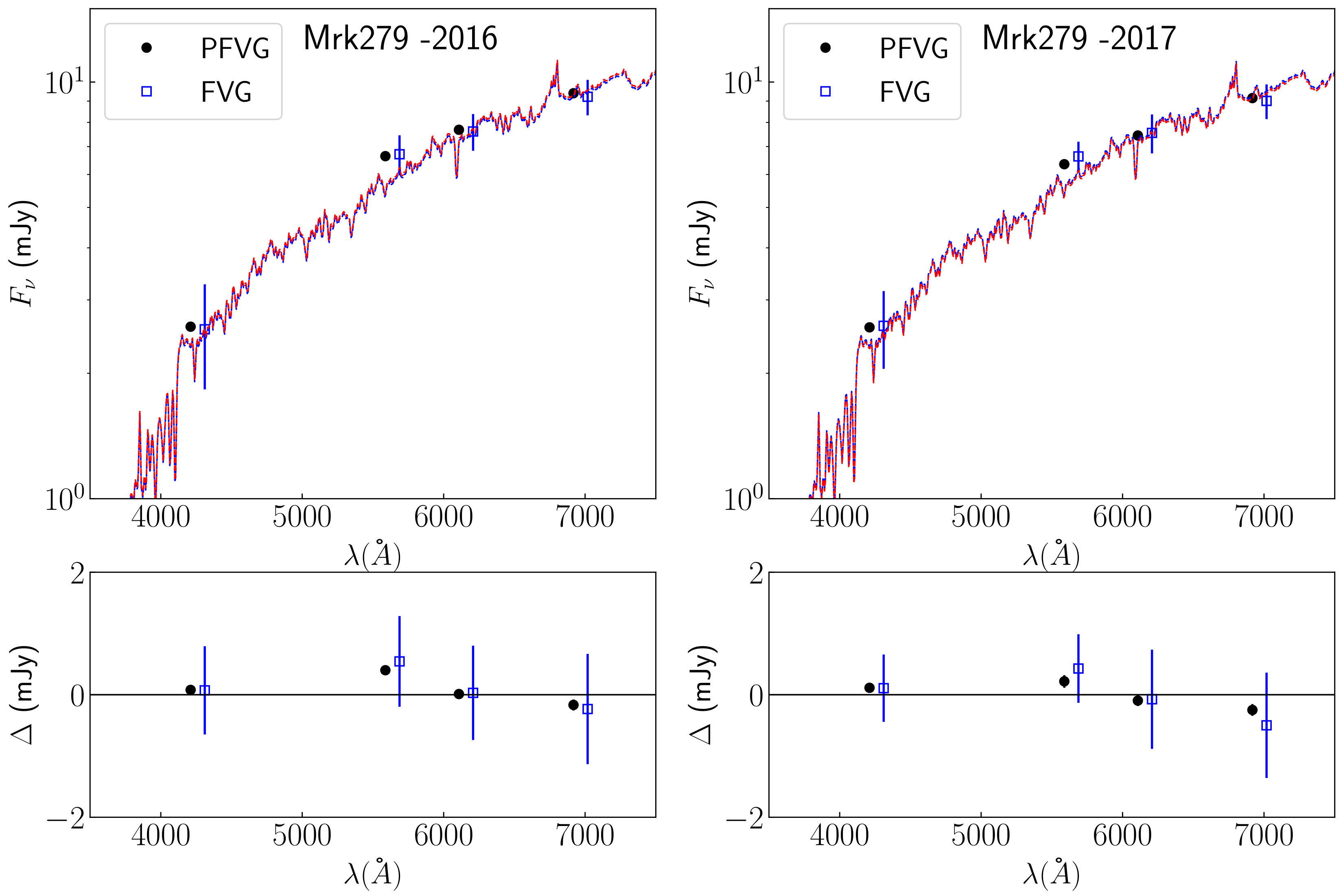}
    \caption{Same as Figure~\ref{fig:AppMrk509}, but for Mrk279 and a redshifted Sa galaxy template of \cite{1996ApJ...467...38K} (red and blue lines).}
    \label{fig:AppMrk279}
\end{figure}

\subsection{Mrk509}
Mrk509 is a nearby ($z = 0.0344$; \citealt{1993AJ....105.1637H}) and bright ($V = 13$ mag) Seyfert 1 galaxy known for its strong variability and characteristic outflows (\citealt{2011A&A...534A..36K}). Through modeling high-resolution Hubble Space Telescope (HST) images, \cite{2009ApJ...697..160B} found a host-galaxy profile consistent with a bulge morphology type. These results are in agreement with the host-galaxy fluxes derived by the traditional FVG analysis presented in \cite{2019MNRAS.490.3936P} (PN+2019). Here we analyze the photometric light curves obtained in two different epochs (2016 and 2017) using the narrow bands with central wavelengths at $4300\pm50$, $5700\pm50$, $6200\pm60$, and $7000\pm60$\,\AA. The quality of the light curves obtained during both epochs is similar, with an average sampling of 1 day and an S/N$\sim150$ for each photometric band. Figure~\ref{fig:AppMrk509} shows the PFVG results along with the FVG obtained by PN+2019. As in PN+2019, we adopted the host-galaxy color given by \cite{2010ApJ...711..461S}. The recovered PFVG distributions are shown in Appendix~\ref{fig:Apx3}. Overall, the results obtained for 2016 and 2017 are consistent, as expected, given the high quality of the light curves. The PFVG is able to recover the galaxy fluxes with a precision of $10$\% compared to a precision of $20$\% obtained by the traditional FVG method. 
These new results lead to a better match with the bulge galaxy template, as can be seen in the fit residuals at the bottom panel of Figure~\ref{fig:AppMrk509}. Deviations from the assumptions of the linear relation between the fluxes (Equation \ref{eq:fall_on_a_line}) due to the accretion disk time delays observed in Mrk509 (PN+2019) are negligible ($< 1$\%).

\subsection{Mrk279}
Mrk279 is a nearby ($z = 0.031$) low-luminosity ($V = 15$ mag) Seyfert 1 galaxy. The FVG analysis by \cite{2019NatAs...3..251C} revealed an S0/Sa host-galaxy morphology, in agreement with the results obtained through modeling high-resolution HST images (\citealt{2002ApJ...569..624P}). The PFVG is able to recover the galaxy fluxes with an exceptional precision of $1$\% compared to the precision of 30\% obtained by the traditional FVG method (Figure~\ref{fig:AppMrk279}). As shown in Appendix~\ref{fig:Apx3}, the average S/N of the light curves is $\sim90$. This is considerably lower than the average S/N$\sim150$ obtained for Mrk509, but consistent with the fact that Mrk279 is about 50\% less luminous. Just as we remarked in the simulations in Section~\ref{sec:pfvgcomp}, we attribute the superior performance of the PFVG to the fact that it explicitly accounts for errors in the measurements.

\begin{figure}
    \centering
    \includegraphics[width=\columnwidth]{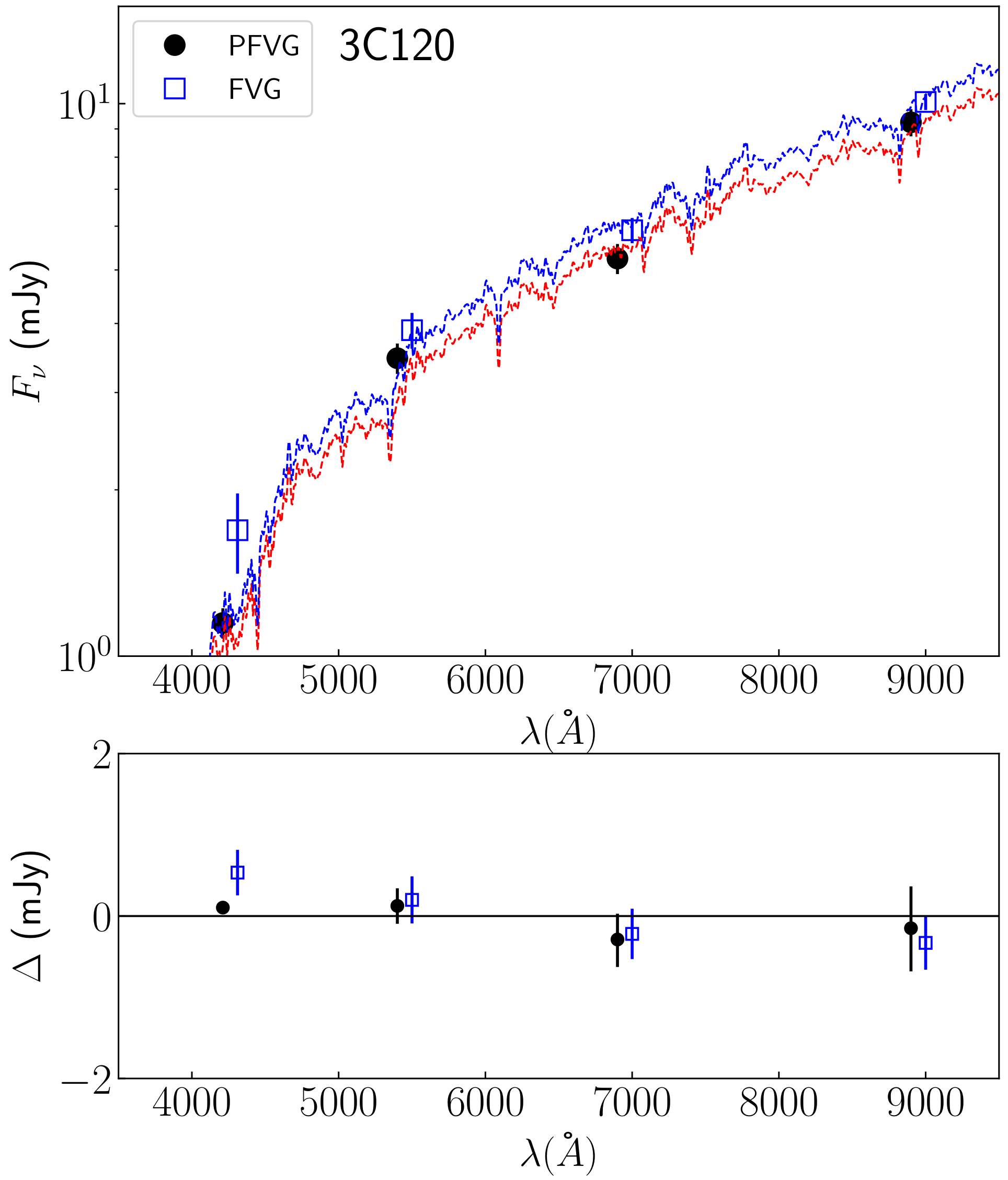}
    \caption{Same as Figure~\ref{fig:AppMrk509}, but for 3C120, and a redshifted S0 galaxy template of \citealt{1996ApJ...467...38K} (red and blue lines).}
    \label{fig:App3C120}
\end{figure}

\subsection{3C120}
3C120 is a nearby Fanaroff-Riley class I radio galaxy at redshift $z = 0.033$ (\citealt{1987ApJ...316..546W}). It has often been the target of several RM campaigns due to its strong variability, with amplitudes of up to $\sim$2 mag on timescales of a few years (\citealt{2010ApJ...711..461S}; \citealt{2012A&A...545A..84P}; \citealt{2012ApJ...755...60G}). More recently, \cite{2018A&A...620A.137R} (R+2018) carried out a PRM campaign in order to study the BLR and the dust torus. R+2018 used the Johnson $B$ and $V$ broadbands to estimate the AGN continuum emission and the $R$ and $I$ broadbands to estimate the fraction of the continuum underneath the H$\alpha$ emission line. The FVG analysis by R+2018 was performed using the $BVI$ bands. The host-galaxy contribution in the $R$ band was not recovered by the FVG due to the contamination of the H$\alpha$ emission line, which is about 14\%\  of the total flux. Moreover, deviations from
the linear relation between the fluxes (Equation \ref{eq:fall_on_a_line}) were noted because the $R$ band contains the delayed H$\alpha$ emission, which does not correlate with the continuum variations traced by the other bands. To overcome these issues, R+2018 fit the spectral energy distribution (SED) with a redshifted S0 (\citealt{2009ApJ...697..160B}) galaxy template and recovered an $R$-band galaxy flux of $5.90\pm0.30$ mJy.

Following R+2018, we adopted a host-galaxy color given by an S0 template (\citealt{1996ApJ...467...38K}). The results are shown in Figure~\ref{fig:App3C120}. Notably, the recovered $R$-band galaxy flux $5.24\pm0.32$ mJy is in full agreement with the SED value reported by R+2018. Previous knowledge of the true color of the host galaxy provides an important advantage for the proposed PFVG method, hence the $R$-band galaxy flux can be directly recovered from the observed light curves. This is possible because the PFVG estimates a density of likely host-galaxy vectors along the assumed host-galaxy line. In this particular case, the method is therefore similar to searching for the best fit in a traditional SED analysis, but with the advantage of using the full information from the observed light curves.

Overall, our results for 3C120 are consistent with those derived from the FVG method, except for the $B$ band, where the PFVG flux is about 1.5 lower than the reported FVG value $1.69\pm0.28$ by R+2018. The PFVG shows a very narrow distribution of fluxes (see Appendix~\ref{fig:Apx4}) with a median value of $1.15$ and a 1$\sigma$ uncertainty of 6\% compared to 16\% obtained by the traditional FVG. The median of the PFVG distribution, however, agrees better with the S0 host-galaxy spectrum shown in Figure~\ref{fig:App3C120}. The better performance of the PFVG is expected because the $B$ band corresponds to the light curve with the lowest S/N = $46$. As shown in the simulations in Section~\ref{sec:III}, if the noise level differs considerably from band to band, then the PFVG method will outperform the FVG, as it weights data items according to their noise measurement.

\section{Summary and conclusions}

We have presented a novel probabilistic flux variation gradient approach to separate the AGN and host-galaxy contributions in the context of photometric reverberation mapping of AGN. We studied its performance using simulated light curves of varying quality, in particular, focusing on the observational characteristics of the next LSST survey. We have demonstrated that by explicitly accounting for error measurements in the light curves, the new probabilistic approach outperforms the traditional FVG in all simulated observational scenarios explored in this work. This leads to recovered galaxy fluxes with a precision lower than 1\% for light curves with a moderate S/N. This represents a significant improvement compared to the 20\% precision obtained by the traditional FVG.

Unlike the traditional FVG, PFVG does not identify just a single line that goes through the given total fluxes, but rather a density of possible lines.
This makes it more robust to the presence of noise and informs us about the uncertainty in our estimates.

Both FVG and PVFG assume that the total flux observations form a line in the space of all filters (see Section~\ref{eq:fall_on_a_line}). Both assume that the only reason that observations may deviate  from this line is due to noise. 
In reality, however, total flux observations may additionally deviate from this line due to time delays that originate in the AGN accretion disk or BLR, for instance. This can lead to galaxy measurements that are biased by less than 5\% in the case of objects with low to intermediate black hole masses. The biases can reach up to 10\% in the case of the highest black hole masses ($\sim 6$ -day  time delay). In view of the overall measurements errors of typically more than 10\% introduced by the absolute calibration of the photometric data, any bias in the PFVG galaxy estimation appears to be negligible. 

We emphasize that the PFVG we presented is not a method for estimating the delay, but a method for separating the host galaxy contribution. Introducing a correction for the time delays requires a different model for the PFVG method, which is beyond the scope of this manuscript. When long delays are suspected, special methods that are based on a cross-correlation analysis or measurement of randomness of the data, for example (e.g., interpolated cross-correlation function (\citealt{1987ApJS...65....1G}), discrete correlation function (\citealt{1988ApJ...333..646E}), von Neumann estimator (\citealt{2017ApJ...844..146C}) could be used to correct for the delay and restore the linear relation between the fluxes on which the PFVG relies.

The application of the PFVG to PRM data sets yielded host-galaxy fluxes that agree better with the host-galaxy morphology types obtained through modeling high-resolution images. As a consequence, the method enables reliable estimates of the pure AGN luminosity, which is necessary to probe theoretical models of the AGN accretion disk, BLR, and dust-torus structure. This is especially important considering the large amount of data that will be provided by future large photometric survey telescopes such as the LSST.

In future endeavors, we would like to address certain assumptions made in the formulation of the proposed PVFG: we currently assume that observational noise is Gaussian distributed, but other choices may be more appropriate, especially in the presence of outliers. Moreover, we assume that no observations are missing and that observations in different filters co-occur (i.e.,~they are observed at the same time instance) so that we can form vectors of them. This limitation prevents the PFVG from combining observations from different surveys because each survey typically has its own observational characteristics (e.g.,~monitoring schedules, cadence).

Finally, an implementation of our method is publicly available through as a \textit{\textup{Julia}} package at \url{https://github.com/HITS-AIN/ProbabilisticFluxVariationGradient.jl}.

\begin{acknowledgements}

The authors gratefully acknowledge the generous and invaluable support of the Klaus Tschira Foundation. This research has made use of the NASA/IPAC Extragalactic Database (NED) which is operated by the Jet Propulsion Laboratory, California Institute of Technology, under contract with the National Aeronautics and Space Administration. This research has made use of the SIMBAD database, operated at CDS, Strasbourg, France. We thank the anonymous referee for his comments and careful review of the manuscript.
      
\end{acknowledgements}

\begin{appendix}

\section{additional tables and figures}

\begin{figure*}
 \includegraphics[width=\columnwidth]{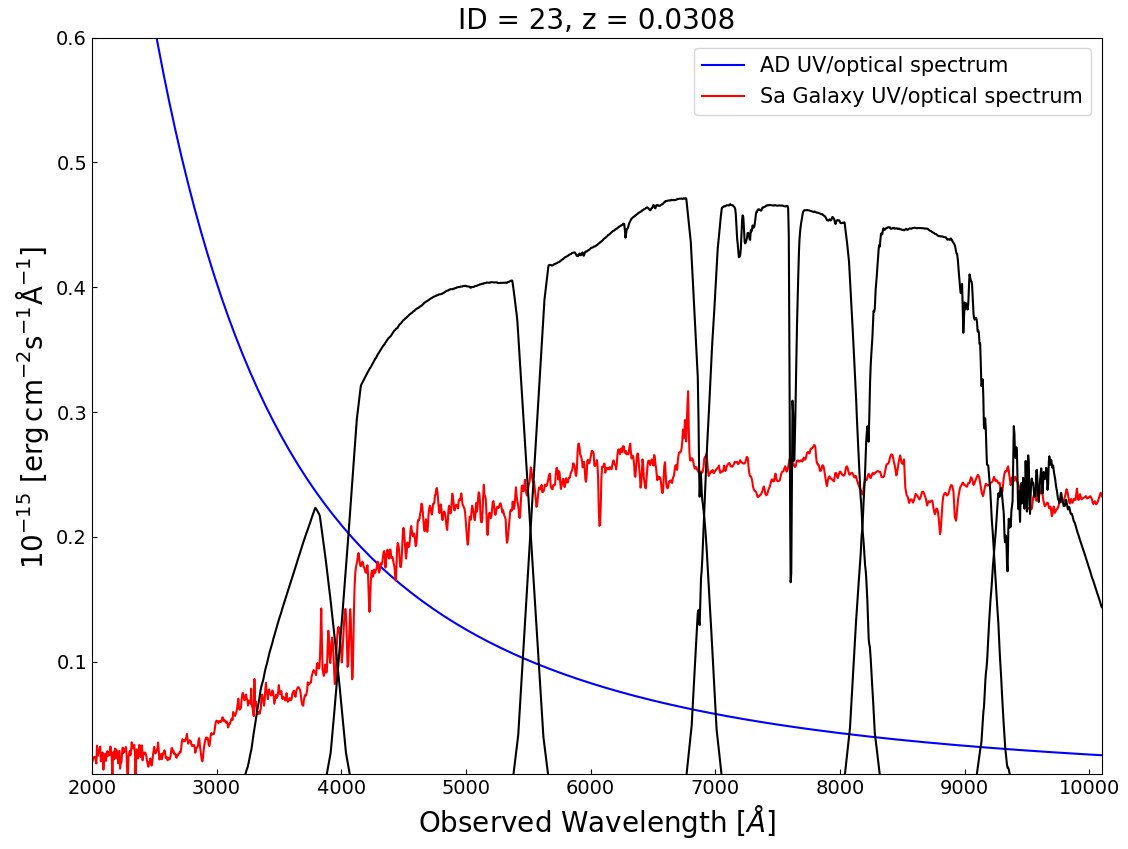}
 \includegraphics[width=\columnwidth]{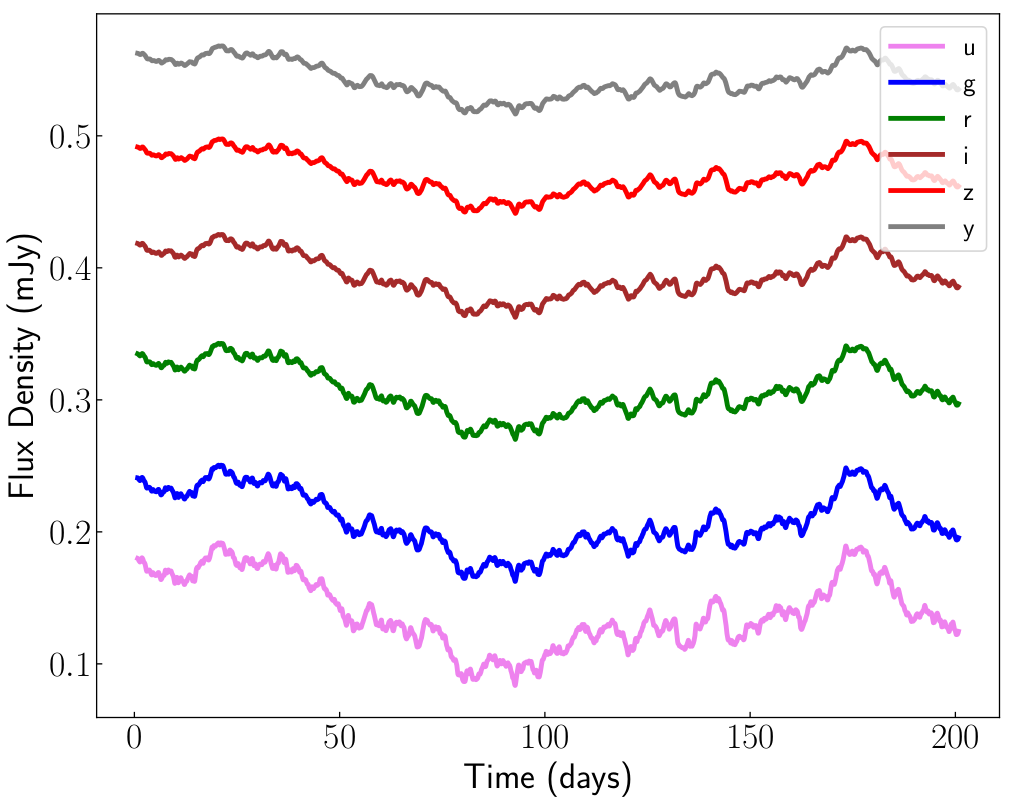}
 \caption{Example of a simulated UV/Optical AGN spectrum (solid blue line, left) along with a Sa-type host-galaxy template (solid red line). The transmission curves of the LSST broadband filters are shown with solid black lines ($ugrizy$; from left to right). In this particular case, the host galaxy contributes about 50\% of the total flux at $5100$\AA\, (rest frame). Flux-calibrated light curves resulting from the convolution between each LSST broadband filter with the AGN and galaxy components (right).  }
 \label{fig:adgalspec}
\end{figure*}

\begingroup
\setlength{\tabcolsep}{4.5pt} 
\renewcommand{\arraystretch}{1.5} 
\begin{table*}
\begin{center}
\caption{PFVG results.}
\label{pfvgval}
\begin{tabular}{@{}ccccccccc}
\hline\hline
Filter & $\rm gal_{0}$  & S/N = 10, $\Delta t=10$ & S/N = 100, $\Delta t=10$ & S/N = 10, $\Delta t=1$ & S/N = 100, $\Delta t=1$ \\
       &                &  CS/HS    & CS/HS    & CS/HS    & CS/HS \\   
\hline
$u$ &$0.897$ &$0.76^{+0.10}_{-0.10}$/$1.85^{+0.63}_{-0.77}$ & $0.89^{+0.01}_{-0.01}$/$1.01^{+0.16}_{-0.18}$ & $0.90^{+0.03}_{-0.03}$/$1.36^{+0.32}_{-0.35}$ & $0.897^{+0.003}_{-0.003}$/$0.879^{+0.038}_{-0.039}$ \\
$g$ &$3.703$ &$3.14^{+0.40}_{-0.43}$/$7.66^{+2.62}_{-3.18}$ & $3.66^{+0.04}_{-0.04}$/$4.14^{+0.67}_{-0.75}$ & $3.71^{+0.1!}_{-0.11}$/$5.63^{+1.31}_{-1.49}$ & $3.704^{+0.010}_{-0.010}$/$3.631^{+0.156}_{-0.162}$ \\
$r$ &$7.589$ &$6.46^{+0.83}_{-0.88}$/$15.62^{+5.35}_{-6.42}$ & $7.50^{+0.07}_{-0.07}$/$8.50^{+1.36}_{-1.54}$ & $7.60^{+0.22}_{-0.22}$/$11.52^{+2.74}_{-2.94}$ & $7.589^{+0.021}_{-0.022}$/$7.433^{+0.317}_{-0.332}$ \\
$i$ &$10.954$ &$9.32^{+1.22}_{-1.28}$/$22.52^{+7.75}_{-9.30}$ & $10.82^{+0.10}_{-0.11}$/$12.24^{+1.97}_{-2.23}$ & $10.96^{+0.31}_{-0.33}$/$16.63^{+3.82}_{-4.24}$ & $10.954^{+0.032}_{-0.032}$/$10.742^{+0.458}_{-0.478}$\\
$z$ &$13.827$ &$11.74^{+1.49}_{-1.61}$/$28.54^{+9.74}_{-11.96}$ & $13.65^{+0.13}_{-0.13}$/$15.53^{+2.55}_{-2.69}$ & $13.84^{+0.39}_{-0.41}$/$20.96^{+4.83}_{-5.52}$ & $13.827^{+0.040}_{-0.040}$/$13.549^{+0.592}_{-0.585}$ \\
$y$ &$16.571$ &$14.08^{+1.84}_{-1.90}$/$34.32^{+11.82}_{-14.33}$ & $16.37^{+0.16}_{-0.16}$/$18.59^{+3.04}_{-3.29}$ & $16.58^{+0.47}_{-0.49}$/$25.12^{+5.85}_{-6.56}$ & $16.570^{+0.047}_{-0.049}$/$16.247^{+0.696}_{-0.719}$ \\
\hline
\end{tabular}
\end{center}
\noindent
{\em Notes:}
$\rm gal_{0}$ corresponds to the true galaxy vector used in the simulations. CS and HS correspond to the results obtained for the AGN combined and higher states, respectively. All fluxes are given in mJy.
\end{table*}
\endgroup

\begin{figure*}
\begin{tabular}{cc}
  \includegraphics[width=85mm]{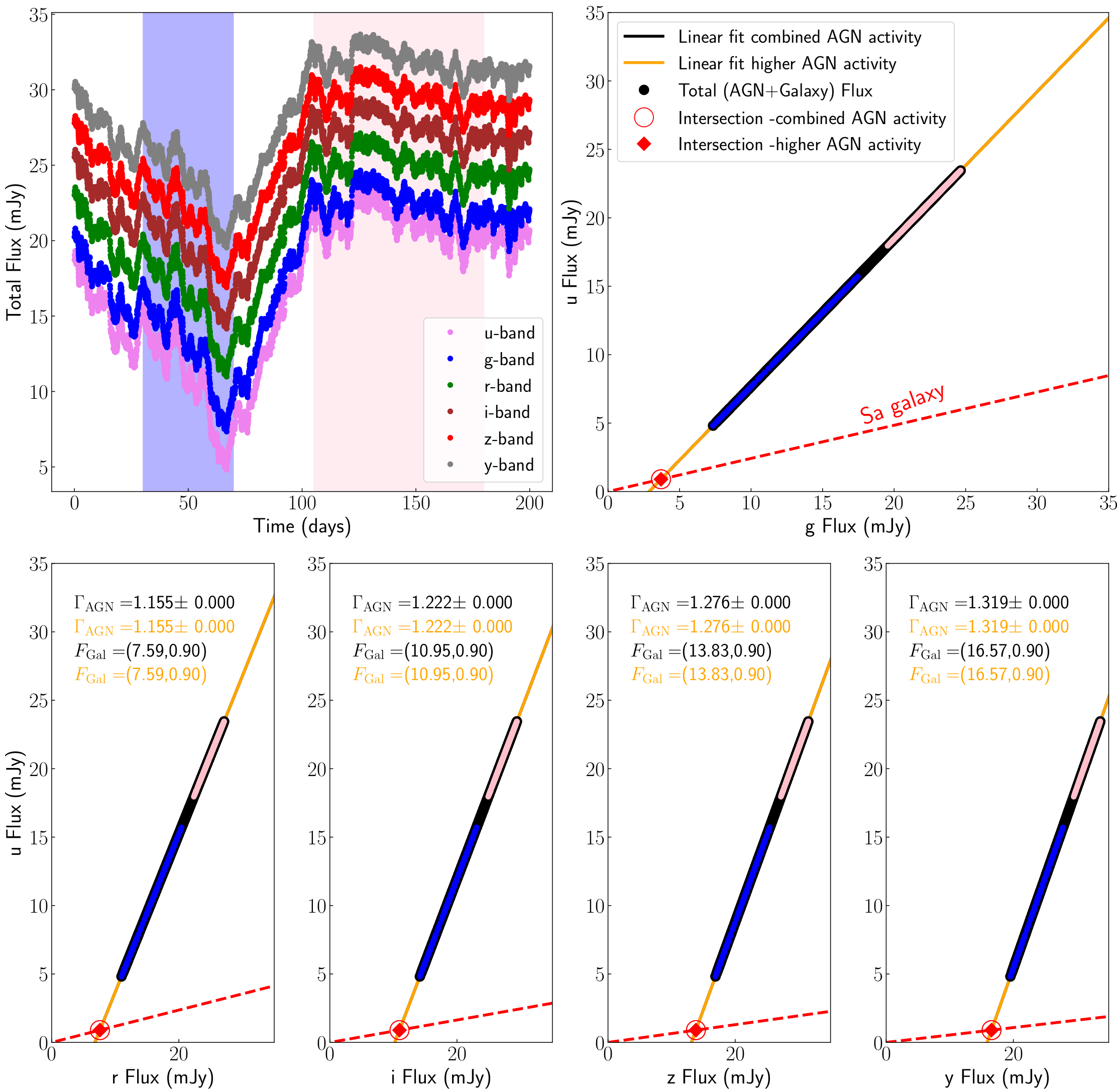} &   \includegraphics[width=85mm]{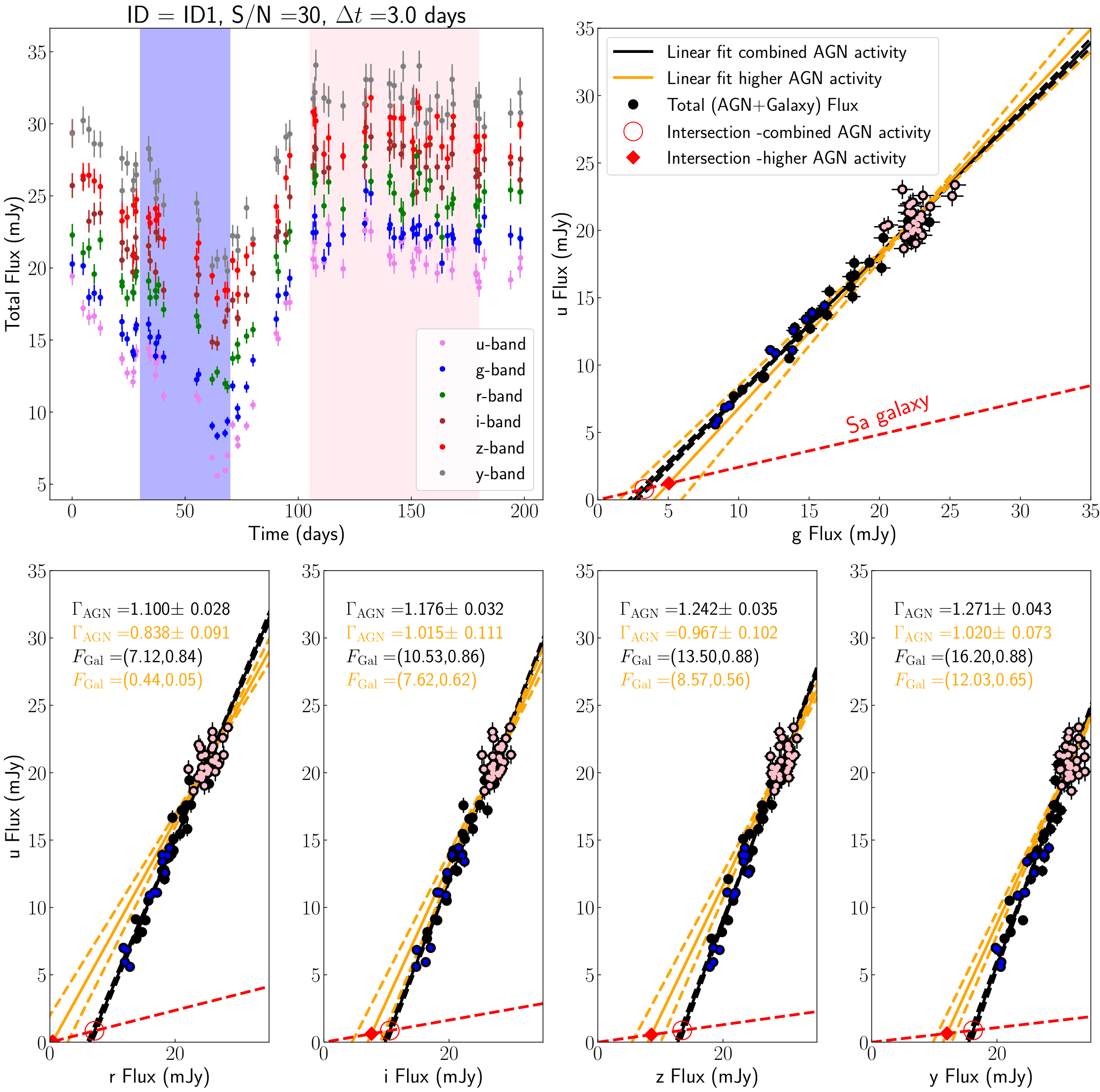} \\
(a)  & (b)  \\[6pt]
 \includegraphics[width=85mm]{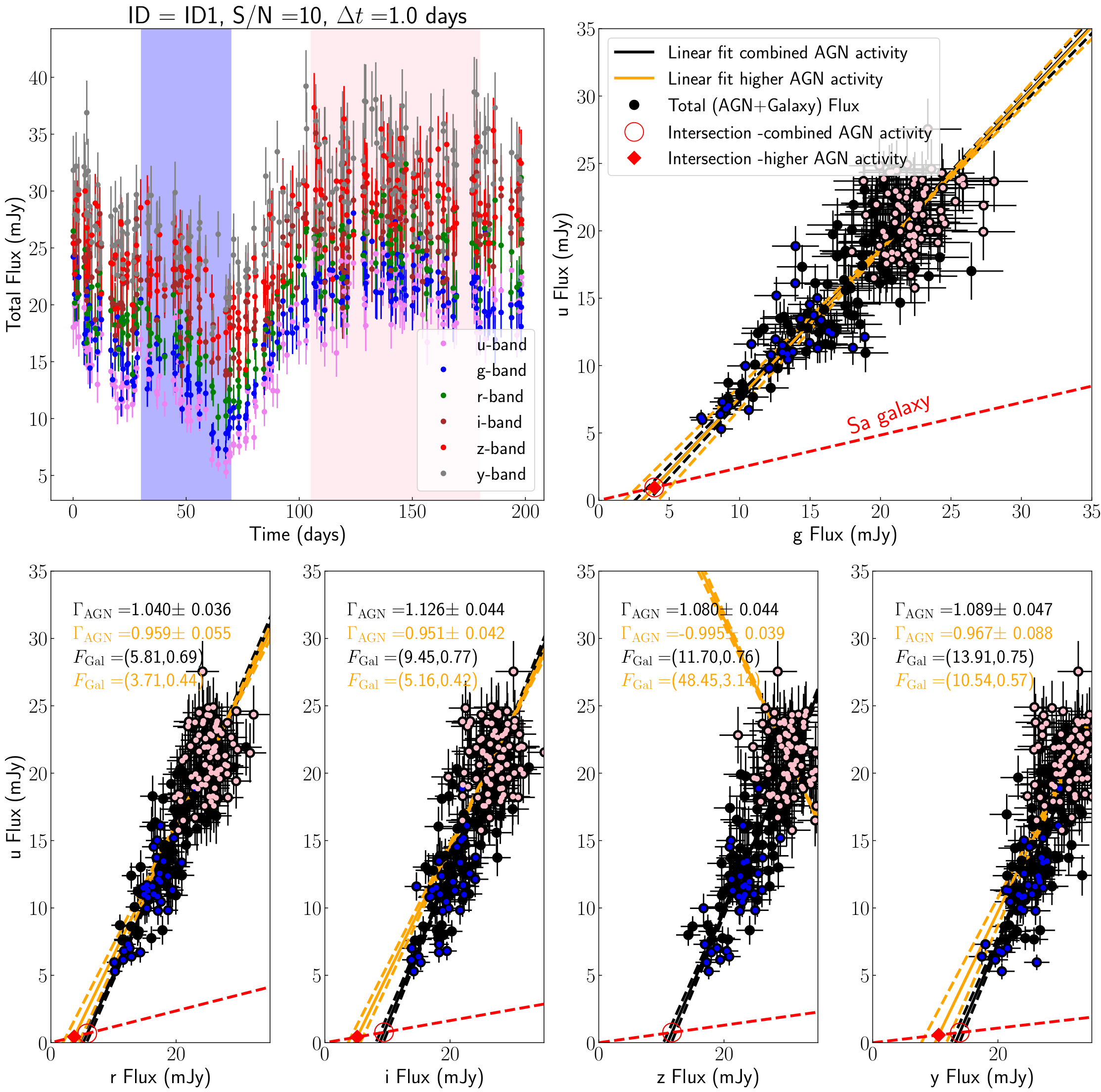} &   \includegraphics[width=85mm]{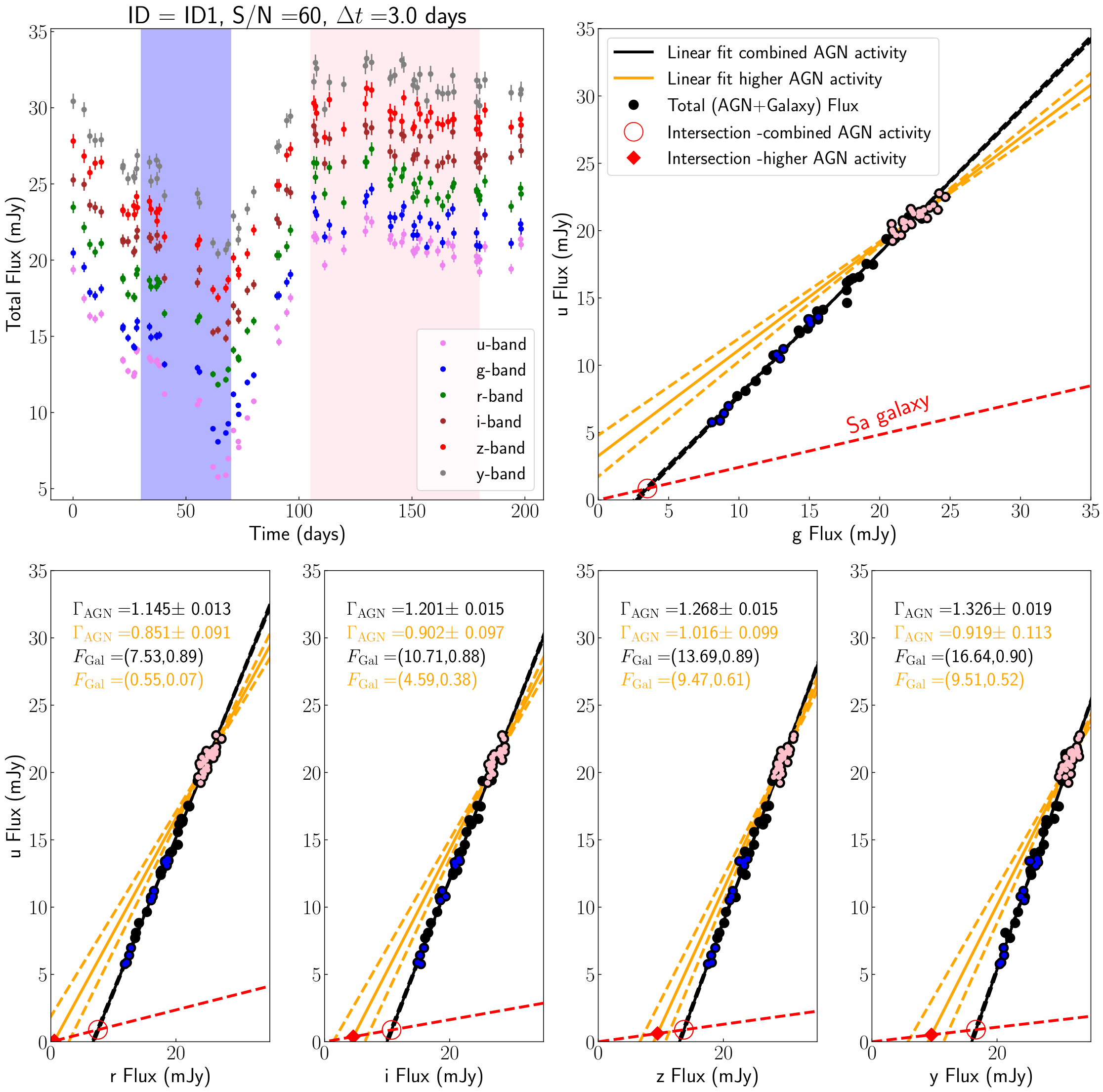} \\
(c)  & (d)  \\[6pt]
\end{tabular}
\caption{FVG for the LSST-simulated light curves from our mock catalog. Panel (a) shows the FVG results obtained for the ideal noiseless light curve with a time sampling of 0.1 days. Panels (b), (c), and (d) show the FVG results for light curves of varying quality.}
\label{fig:Apx1}
\end{figure*}

\begin{figure*}
\begin{tabular}{cc}
  \includegraphics[width=85mm]{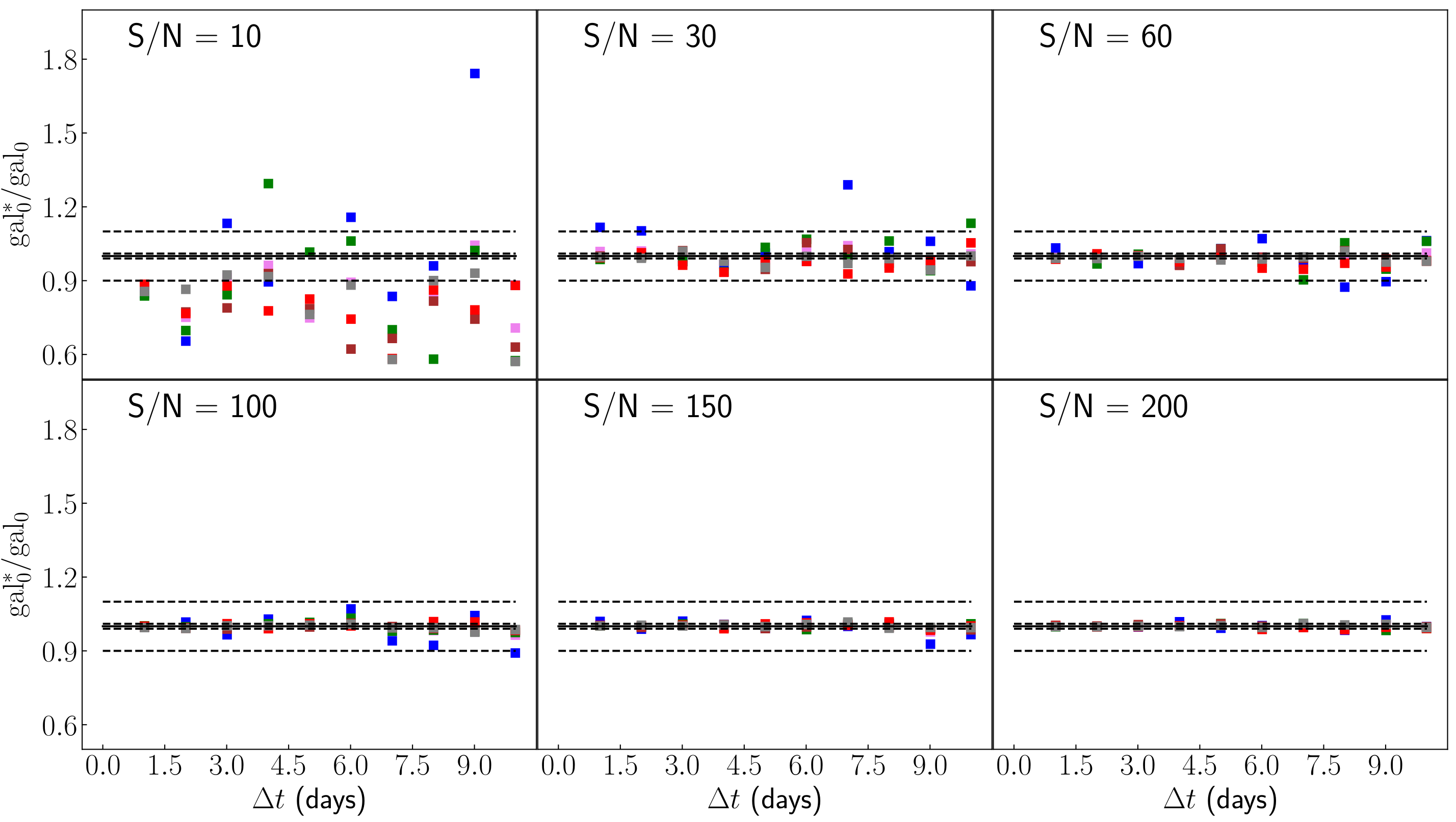} &   \includegraphics[width=85mm]{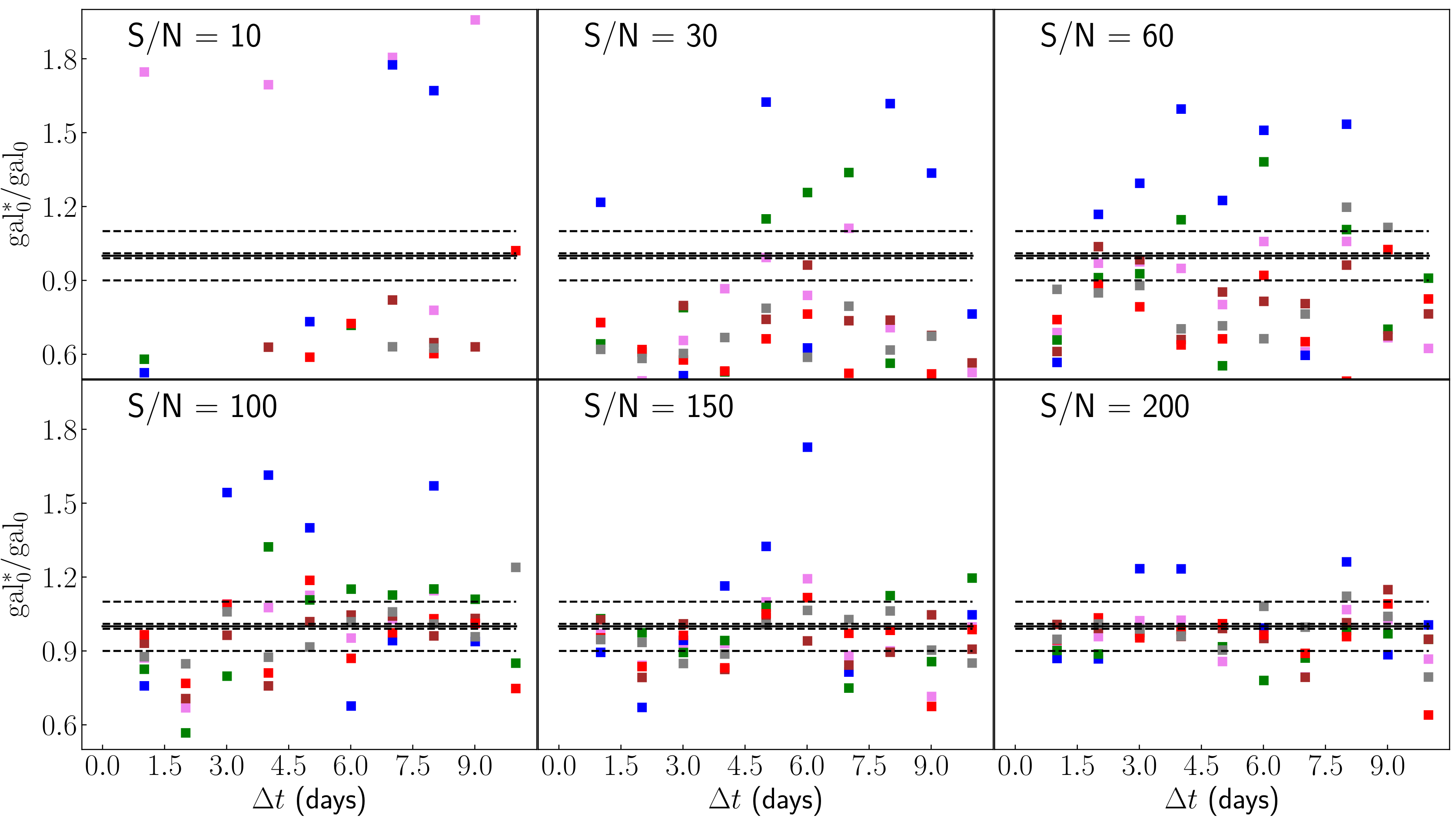} \\
(a)  & (b)  \\[6pt]
 \includegraphics[width=85mm]{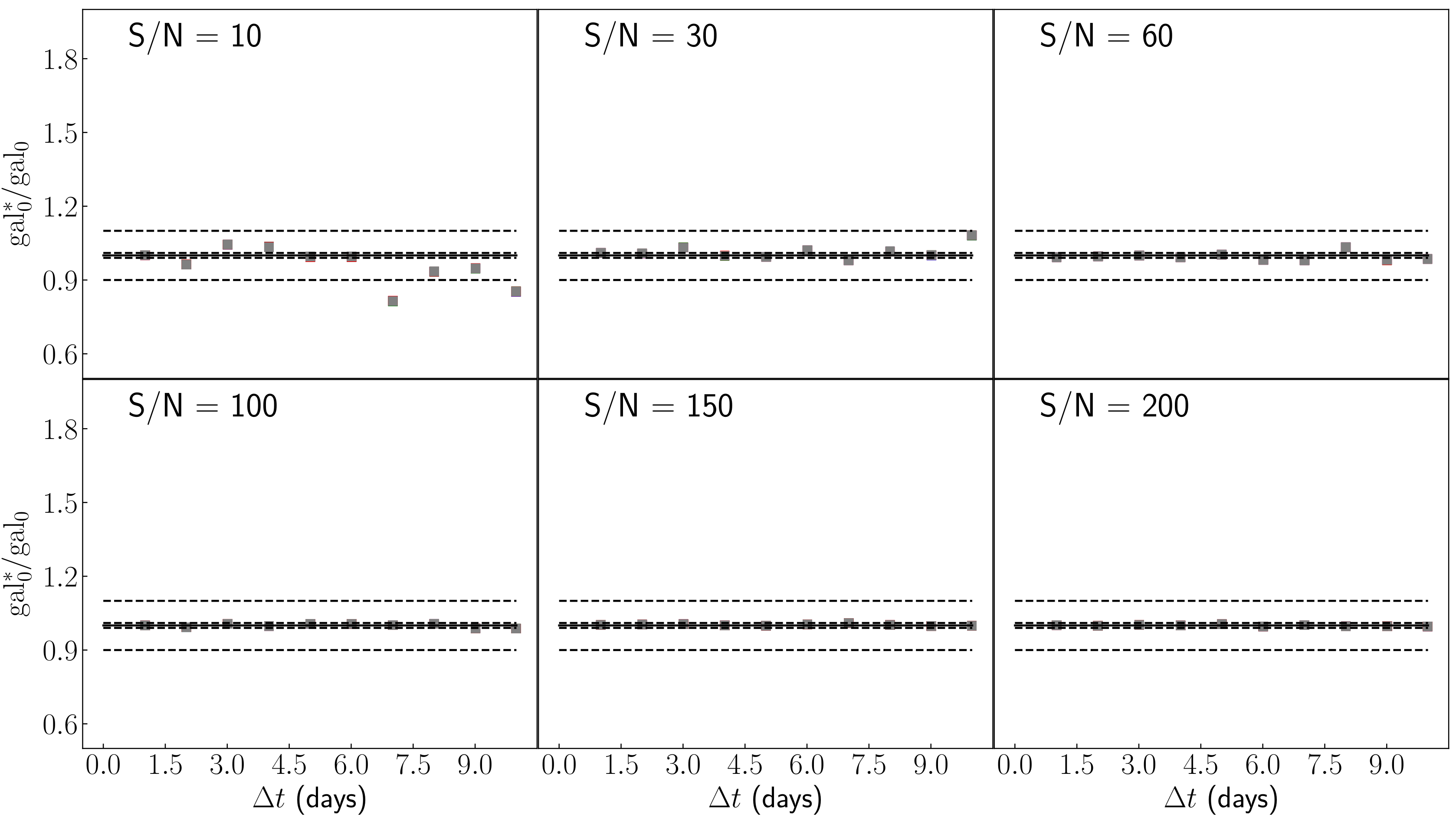} &   \includegraphics[width=85mm]{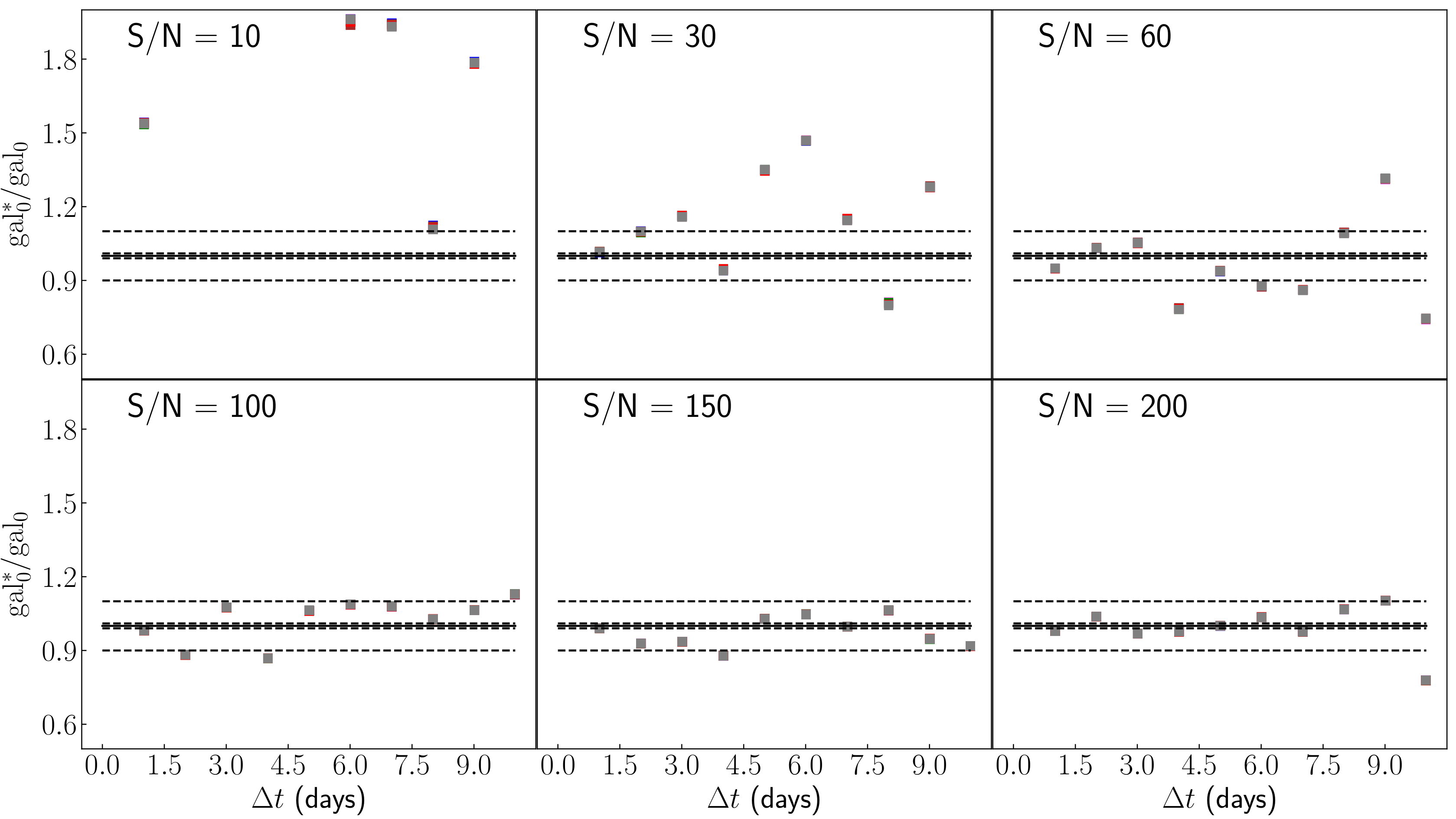} \\
(c)  & (d)  \\[6pt]
\end{tabular}
\caption{Recovered FVG (top) and PFVG (bottom) distributions of host-galaxy fluxes ($\rm gal^{*}_{0}$) for different S/N and time sampling. The left and right panels show $\rm gal^{*}_{0}$ obtained for the CS and HS, respectively.}
\label{fig:Apx2}
\end{figure*}



\begin{figure*}
\begin{tabular}{cc}
  \includegraphics[width=85mm]{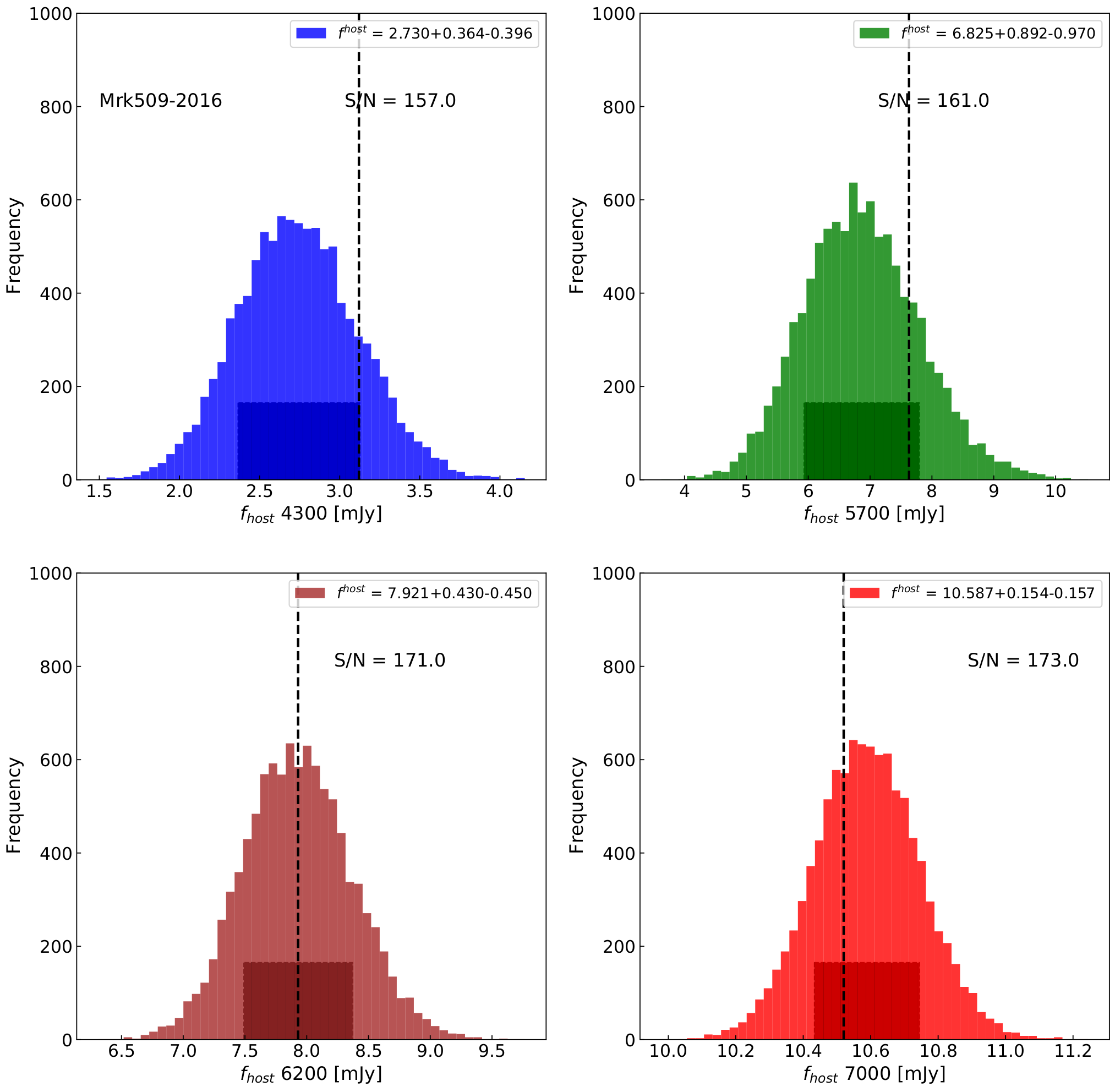} &   \includegraphics[width=85mm]{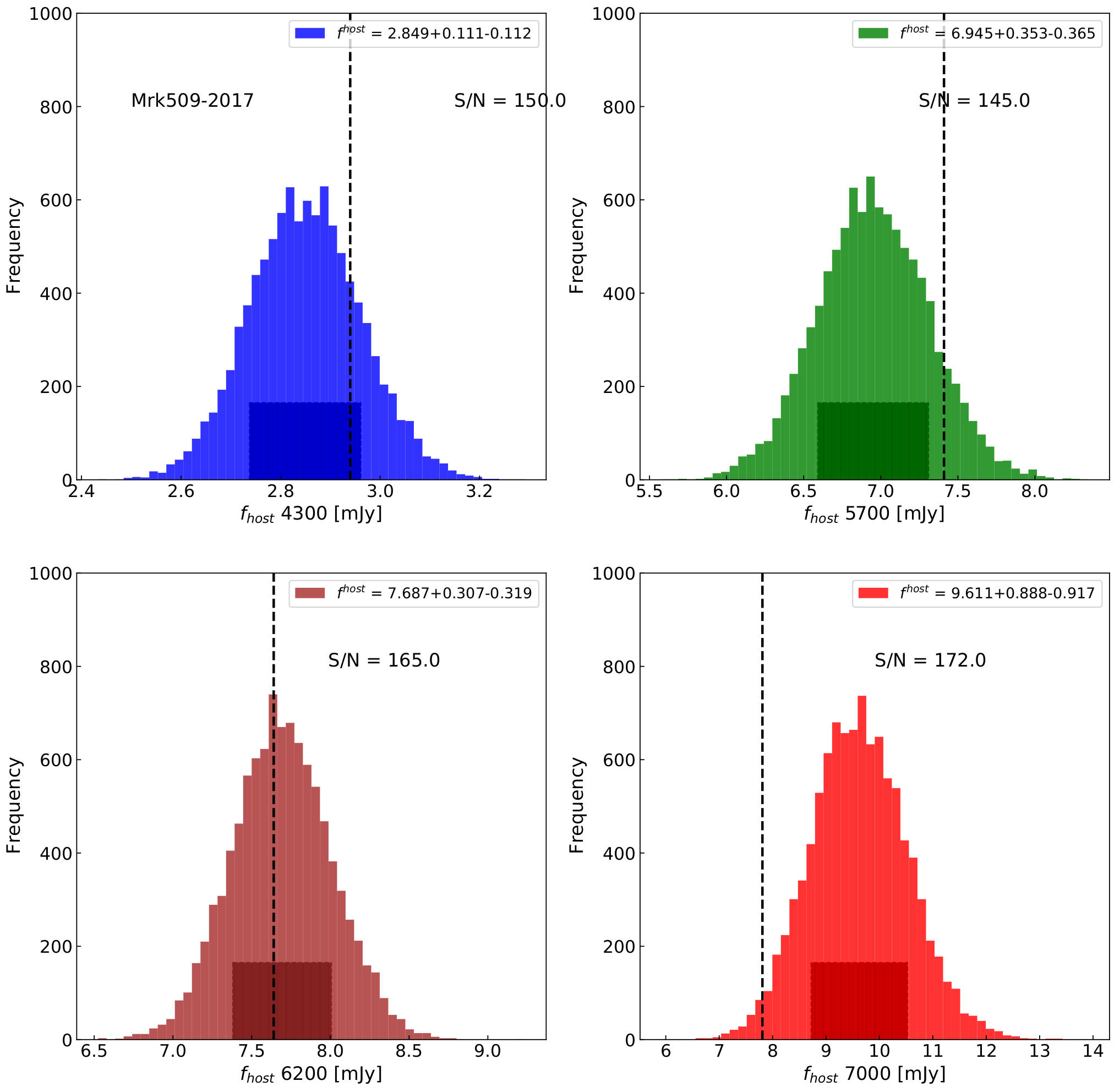} \\
(a)  & (b)  \\[6pt]
 \includegraphics[width=85mm]{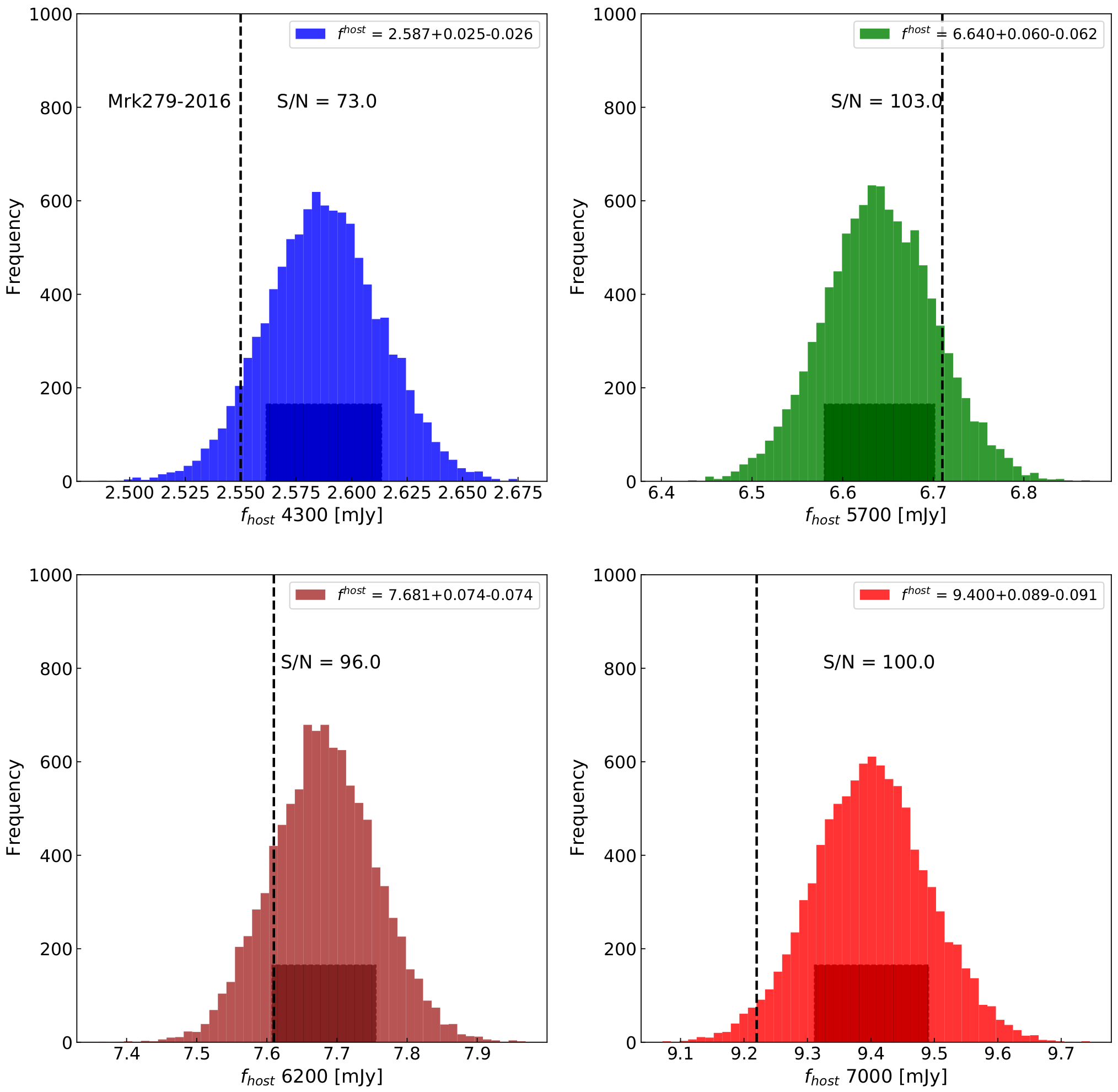} &   \includegraphics[width=85mm]{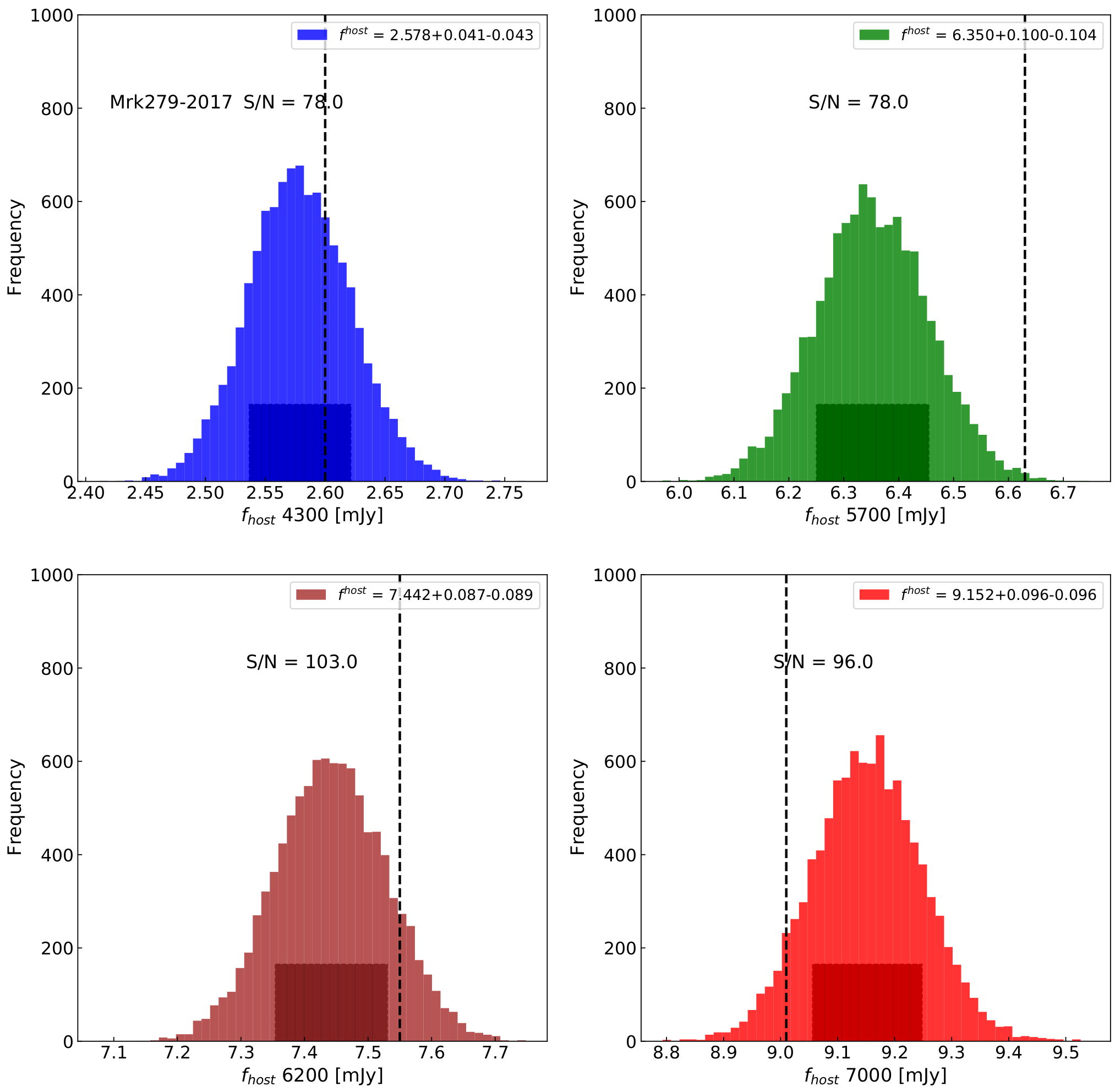} \\
(c)  & (d)  \\[6pt]
\end{tabular}
\caption{Recovered PFVG distributions of host-galaxy fluxes for objects Mrk509 (top panels) and Mrk279 (bottom panels). The vertical dotted line marks the galaxy flux obtained by the FVG method. The dark shaded area marks the 68\% confidence range we used to estimate the 1$\sigma$ uncertainty around the median of the distribution.}
\label{fig:Apx3}
\end{figure*}

\begin{figure*}
\begin{tabular}{cc}
  \includegraphics[width=85mm]{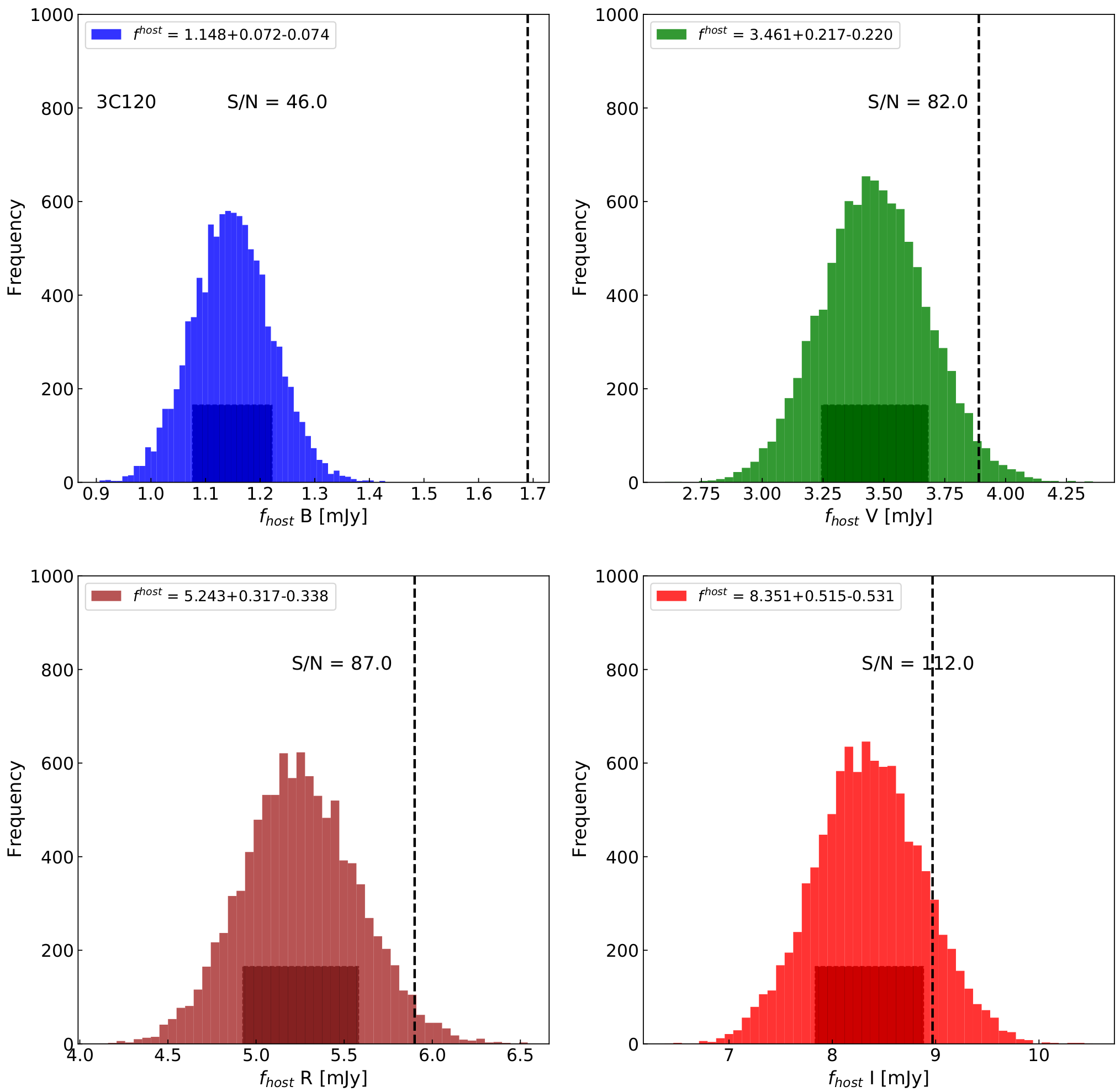}
\end{tabular}
\caption{Same as Figure~\ref{fig:Apx3}, but for 3C120.}
\label{fig:Apx4}
\end{figure*}

\end{appendix}

\end{document}